\documentclass[12pt]{article}
\usepackage{epsf,cite,a4wide,latexsym}
\newcommand{\be}{\begin{equation}}
\newcommand{\ee}{\end{equation}}
\newcommand{\bea}{\begin{eqnarray}}
\newcommand{\eea}{\end{eqnarray}}

\newcommand{\Tr}{{\rm Tr}}
\newcommand{\N} {{\rm N}}
\newcommand{\Id}{1\!\!1}
\newcommand{\half}{\frac{1}{2}}

\newcommand{\basispl}{
   \put(-.5,-.5){\line(1,0){1}}
   \put(.5,-.5){\line(0,1){1}}
   \put(.5,.5){\line(-1,0){1}}
   \put(-.5,.5){\line(0,-1){1}}
                     }

\newcommand{\basisar}{
   \put(0,-.5){\vector(1,0){0}}
   \put(.5,0){\vector(0,1){0}}
   \put(0,.5){\vector(-1,0){0}}
   \put(-.5,0){\vector(0,-1){0}}
                     }

\newcommand{\clover}{\setlength{\unitlength}{.5cm}\raisebox{-.5cm}{
   \begin{picture}(2.4,2.4)(-1.2,-1.2)
   \multiput(-1.2,-1.2)(1.2,1.2){2}{\begin{picture}(1.2,1.2)(-.6,-.6)
   \basispl\basisar\end{picture}}
   \multiput(-1.2,0)(1.2,-1.2){2}{\begin{picture}(1.2,1.2)(-.6,-.6)
   \basispl\basisar\end{picture}}
   \put(-.1,-.1){\circle*{.2}}
   \put(-.1,.1){\circle*{.2}}
   \put(.1,-.1){\circle*{.2}}
   \put(.1,.1){\circle*{.2}}
   \end{picture}}}

\begin{document}
\vskip-1cm
\hfill MIT-CTP-2972
\vskip5mm

\begin{center}
{\LARGE{\bf{Numerical analysis of fractional}}}\\
\vspace{0.3 cm}
{\LARGE{\bf{charge solutions on the torus}}}\\
\vspace{0.3 cm}

\vspace{1.0 cm}

{\Large \'Alvaro Montero   \\
\vspace{-0.4 cm}
Center for Theoretical Physics \\
\vspace{-0.4 cm}
Laboratory for Nuclear Science 
and Department of Physics \\
\vspace{-0.4 cm}
Massachusetts Institute of Technology \\ 
\vspace{-0.4 cm}
Cambridge, Massachusetts 02139 \\
\vspace{-0.4 cm}
USA \\}
\vspace{0.4 cm}
{\Large e-mail: montero@mitlns.mit.edu}

\end{center}

\vspace*{5mm}{\narrower\narrower{\noindent
\begin{center}
{\bf ABSTRACT}
\end{center}
We study by numerical methods
a particular kind of SU(N) Yang-Mills solutions
of the Euclidean equations of motion which appear on the torus
when twisted boundary conditions are imposed. These are 
instanton-like configurations with the peculiarity of having 
fractional topological charge. We focus on those solutions with
minimal non-trivial action $S=8\pi^2/\N$ and extract their properties
in a few different cases, paying special attention to the 
$N \rightarrow \infty$ limit.}}

\newpage

\section{Introduction}
Quite some time ago 't Hooft \cite{torus} pointed out some 
very special features which arise while formulating gauge theories 
on a torus (a review can be found in reference \cite{tonrev}). 
They are related to the freedom in choosing boundary 
conditions for the gauge potential:
since only local, gauge invariant quantities are required to be periodic,
periodicity of the gauge potential has to be satisfied only up to a gauge 
transformation. Under displacement of the gauge potential 
by a torus period $l_\nu$ in direction $\hat{\nu}$ 
\begin{eqnarray}
{\bf A}_{\mu}(x+l_{\nu}\hat{\nu})=
{\bf \Omega}_{\nu}(x) {\bf A}_{\mu}(x) {\bf \Omega}_{\nu}^{\dagger}(x)
-\imath{\bf \Omega}_{\nu}(x)\partial_{\mu}{\bf \Omega}_{\nu}^{\dagger}(x)\ ,\ \mu=1,\cdots,4,
\end{eqnarray}
with  ${\bf \Omega}_{\nu}(x_{\mu\ne\nu})$
SU(N) matrices, also known as the twist matrices. The choice of such
matrices is arbitrary  up to a consistency condition derived
from the requirement of single-valuedness of the gauge potential.
Built in two ways from ${\bf A}_{\mu}(x_{\nu},x_{\rho})$ 
(through ${\bf A}_{\mu}(x_{\nu}\!+\! {l_{\nu}},x_{\rho})$ and
 ${\bf A}_{\mu}(x_{\nu},x_{\rho}\!+\!{l_{\rho}})$)   single-valuedness for 
${\bf A}_{\mu}(x_{\nu}\!+\!{l_{\nu}}, x_{\rho}\!+\!{l_{\rho}})$ implies
\bea
 {\bf \Omega}_{\mu}(x_{\nu} + {l_{\nu}} ) {\bf \Omega}_{\nu}(x_{\mu} ) 
=  {\bf \Omega}_{\nu}(x_{\mu} + {l_{\mu}} ) {\bf \Omega}_{\mu}(x_{\nu}  ) 
 \ {\rm Exp} \left(-\frac{ \imath 2 \pi n_{\mu \nu} }{\N} \right)
\eea
with $n_{\mu\nu}$ a gauge invariant antisymmetric tensor of integers, 
defined modulo N and independent of $x$ (this twist factor is 
allowed due to the invariance of ${\bf A}_\mu$ under 
a gauge transformation with an element of the center $Z\!\!\!Z_\N$ 
of  SU(N)).
Indeed, the actual choice of the twist matrices is irrelevant and
only the consistency conditions given by $n_{\mu\nu}$ matter. 
Whenever $n_{\mu\nu}\!\neq\! 0$ (mod N), for some $\mu$, $\nu$,
we say the boundary conditions are twisted.
The twist is reflected in a gauge invariant way through the 
non-trivial periodicity of the Polyakov loops, defined on the torus as
\be
{\cal L}_{\mu}(x)=\frac{1}{\N}\Tr\ ({{\bf L}_\mu})=\frac{1}{\N} 
\Tr  \left ( {\rm Texp}\left\{\imath\int_{\gamma_{\mu(x,x')}} {\bf A}_{\nu}dx^{\nu}\right\}
\hspace{0.1 cm} {\bf \Omega}_{\mu}(x') \hspace{0.1 cm}
{\rm Texp}\left\{\imath\int_{\gamma_{\mu(x',x)}} {\bf A}_{\nu}dx^{\nu}\right\}
\right ) \label{eq:pol}
\ee
with $\gamma_{\mu(a,b)}$ a straight line in the positive $\mu$ direction 
starting at $a$ and ending at $b$ and $x'$ the border of the torus patch. 
Periodicity holds only up to the twist factors,
 i.e.
\be
{\cal L}_\mu(x+l_\nu \hat{\nu})= {\rm Exp} \left(-\frac{\imath 2 \pi n_{\mu
\nu} }{\N} \right) {\cal L}_\mu(x).  \label{eq:bcpl}
\ee

With twisted boundary conditions the topological 
charge Q is no longer necessarily an integer:
\begin{equation}
Q = \frac{1}{16 \pi^2} \int \Tr \left( {\bf F}_{\mu \nu} \tilde{{\bf F}}_{\mu \nu} \right) 
d_4x = \nu - \frac{\kappa}{\N}  , \ \ {\rm with } \ \ \nu, \kappa \in Z\!\!\!Z,
\end{equation}
\noindent
$\kappa$ is associated to the matrix of integers $(n_{\mu \nu})$ through
\begin{equation}
 \kappa = \frac{1}{4} n_{\mu \nu} \tilde{n}_{\mu \nu} = \vec{k} \cdot \vec{m} ,
\end{equation}
\noindent
with $k_i=n_{0i}$, $n_{ij}=\epsilon_{ijk}m_k$. The mathematical proof of this
relation for the topological charge can be found in \cite{vbaal}. Then $Q$ is  
fractional and proportional to $1/\N$ whenever 
$\vec{k} \cdot \vec{m} \ne 0 \ {\rm modulo\  \N}$ (non-orthogonal twist). 
This implies, through Schwarz-inequality, that the action of any 
configuration is bounded from below  by
\begin{equation}
S = \frac{1}{2} \int \Tr \left( {\bf F}_{\mu \nu} {\bf F}_{\mu \nu} \right) d_4x  
\ge 8 \pi^2 |Q|= 8 \pi^2 \left|\nu-\frac{\kappa}{\N}\right|\quad  
\end{equation}
\noindent
with the bound  saturated for self or anti-self dual configurations 
(${\bf F}_{\mu \nu}= \pm \tilde{{\bf F}}_{\mu \nu}$). 
It is clear that whenever $\kappa \ne 0 \ ({\rm modulo\  \N})$
there is an obstruction for zero-action configurations.
Minimal action is attained in such cases if $|\nu-\kappa/\N|=1/\N$ with $S=8\pi^2/$N. 
These are in fact the kind of solutions we will describe in this paper.

Some of these fractional charge solutions have already been found either 
analytically or numerically. 't Hooft has explicitly
constructed non-abelian solutions with constant field strength which turn
out to be (anti-)self-dual whenever the sides of the torus satisfy certain
relations (see \cite{thoof2} for details or the appendix at the end of this paper).
There are also a few numerical studies of solutions with non constant field
strength. The first one is presented in reference \cite{Cool1} and it is 
obtained there the fractional charge solution with $|Q|=1/2$ and $S=4\pi^2$
for the $SU(2)$ group, on a $L^3 \times T$ 
torus with $T \gg L$ and satisfying twisted boundary conditions given by the 
twist vectors $\vec m = (1,1,1)$ and $ \vec k = (1,1,1) $. A full parametrization
of the field strength ${\bf F}_{\mu \nu}$, and of the gauge field ${\bf A}_{\mu}$
for this solution is presented in reference \cite{Cool2}. Another SU(2) solution
is presented in \cite{TA3}, in this case the fractional charge solution with
$|Q|=1/2$ and $S=4\pi^2$, on a torus $L^2 \times T^2$ with $T \gg L$ and
satisfying the appropriate boundary conditions to have the properties of a
vortex. The same kind of solution for the SU(3) group, with $|Q|=1/3$ and
$S=8\pi^2/3$, is presented in reference \cite{A1}, and the generalization 
to SU(N) group with $N>3$ can be found in \cite{A2}. In this article we present
a numerical study of SU(N) solutions, with charge $|Q|=1/N$ and action $S=8\pi^2/N$,
and living on a $L^3 \times T$ torus with $T \gg L$. Some preliminary
results have been presented in \cite{TAMC}.

These configurations are interesting by itself from a mathematical point of view, 
and  physically interesting for their possible relevance in low energy phenomena 
like the confinement property or the breaking of the  chiral symmetry.
As has been pointed out in \cite{thoof2} these solutions may play a role in 
the properties of the theory in the limit of large number of colors, N. 
One of the arguments to question the contribution of instantons to 
long-distance phenomena as confinement is based
on the large N expansion \cite{largeN}. Any instanton mediated interaction 
is suppressed by the semi-classical factor exp$(-8\pi^2/g^2)$,
since the large N limit is achieved while keeping $g^2$N fixed, integer charge
instantons are (at least in the dilute gas picture) naively suppressed by exp(-N). 
The argument no longer holds for twisted instantons with action $8\pi^2/$N.
Another interesting point is the possible relation between the center vortex picture
of confinement, proposed in \cite{vortex} and now being investigated 
\cite{green,lang,cher,forcrand,kovacs}, and fractional charge solutions. As have been 
pointed out in \cite{TA3,A1,A2}, it is possible to built vortex configurations
in $R^4$ from solutions of the Yang Mills equations of motion in $T^4$. 
We also want to mention the model of confinement based in fractional charge objects
presented in reference \cite{TP}, and some favourable results shown in \cite{TA1}. 

The paper is structured as follows. 
In section two the numerical method to obtain the solutions
is briefly described.  We will be interested in solutions living
on a $L^3 \times T$ torus with $T \gg L$ which,
in the limit $T\rightarrow \infty$, represent vacuum to vacuum
tunneling. The analysis will be restricted to spatial twist $ \vec m=(1,1,1)$. 
Section three presents a detailed analysis of these solutions.
Our conclusions are presented in section four. Finally, we include an appendix with
the analytic solutions obtained through 't Hooft construction. 
Their relation with the cases we have studied is discussed through the text.

\section{Numerical minimization of the action}
To generate numerically the minimal action configurations we follow
the method which has allowed to successfully extract these kind of 
solutions for other sizes of the torus and values of the number of
colors in references \cite{Cool1,Cool2,TA3,A1,A2,TAMC}.  
We use the standard discretization of Yang-Mills theories on the lattice 
\cite{Wilson}.  We work on $N_s^3 \times N_t$ lattices, $N_t \gg N_s$, 
with variables defined 
on each link of the lattice taking values on N $\times$ N unitary 
matrices $\hat{{\bf U}}_\mu(n)$.
The lattice action used is the Wilson action,
\be
S_W=\sum_{n,\mu,\nu} 
\Tr\left(1-\hat{{\bf U}}_\mu(n)\hat{{\bf U}}_\nu(n+\hat{\mu})\hat{{\bf U}}_\mu^{\dagger}(n+\hat{\nu})
\hat{{\bf U}}_\nu^{\dagger}(n)\right)\quad,
\ee
where $\mu$ and $\nu$ specify directions (from 1 to 4) and $\hat{\mu}$ , 
$\hat{\nu}$ are unit vectors along the corresponding direction, $n_\mu=1,...,N_\mu$.

The link variables $\hat{{\bf U}}_\mu(n)$ satisfy the (twisted) boundary conditions,
\be
\hat{{\bf U}}_\mu(n+N_{\nu}\ \hat{\nu})= {\bf \Omega}_{\nu}(n)\ \hat{{\bf U}}_\mu(n)\
{\bf \Omega}_\nu^{\dagger}(n+\hat{\mu}))\quad,
\ee
where $N_4\!=\!N_t$, $N_i\!=\!N_s,\ i=1,2,3$ and ${\bf \Omega}_{\mu}$ are 
the twist matrices with consistency condition,
\be
{\bf \Omega}_{\mu}(n+N_{\nu}\hat{\nu})\ {\bf \Omega}_{\nu}(n)=
{\bf \Omega}_{\nu}(n+N_{\mu}\hat{\mu})\ {\bf \Omega}_{\mu}(n)\ 
{\rm Exp}(-2\pi \imath n_{\mu\nu}/\N)\quad.
\ee
It is possible to make a  change of variables
\be
 {\bf U}_\mu(N_\mu,n_\nu)= \hat{{\bf U}}_\mu(N_\mu,n_\nu) {\bf \Omega}_\mu(n_\nu)  \hspace{0.5 cm}
 {\bf U}_\mu(n_\mu \neq N_\mu,n_\nu)= \hat{{\bf U}}_\mu(n_\mu \neq N_\mu,n_\nu)   \label{change}
\ee
such that the new link variables are strictly periodic.
In terms of the new links
\be
S_W=\sum_{n,\mu,\nu} 
\Tr\left(1-Z_{\mu\nu}^*(n)\ {\bf U}_\mu(n)\ {\bf U}_\nu(n+\hat{\mu})\
{\bf U}_\mu^{\dagger}(n+\hat{\nu})\ {\bf U}_\nu^{\dagger}(n)\right)\quad,
\ee
where $Z_{\mu\nu}(n)\in Z\!\!\!Z_\N$ take the values:
$Z_{\mu\nu}(n)=1$ for all plaquettes except the one at the top-right corner in the
$(\mu,\nu)$ plane which is equal to ${\rm Exp}(-2\pi in_{\mu\nu}/\N)$.

The strategy to obtain the solution is minimize the lattice action
with respect to the variable ${\bf U}_{\mu}(n)$ (this minimization procedure 
is usually known as cooling).
We use the Cabibbo-Marinari-Okawa algorithm \cite{su2trick} 
in which each link variable is updated in the way:
${\bf U}_\mu(n) \rightarrow {\bf A}  {\bf U}_\mu(n)$,
where ${\bf A}$ is a SU(N) matrix built from a SU(2) matrix ${\bf a}$ which is embedded
into one of the $N(N-1)/2$ subgroups of SU(N). Once we obtain the matrix ${\bf a}$
minimizing the new action $S_W ({\bf A}  {\bf U}_\mu(n))$, we
update the link variable ${\bf U}_\mu(n)$, and repeat the procedure for all the $N(N-1)/2$
subgroups of SU(N) and for all lattice sites. This constitutes one cooling sweep. 
We iterate this procedure up to we obtain that the Wilson action is stable with a
given precision (in this work, the eight relevant digit) and close to the value 
of the expected continuum action: $S=8\pi^2/N$.

\section{The solutions}

As mentioned in the introduction, we are interested in solutions with minimal
non-trivial action,
\be
S = 8 \pi^2 |Q| = \frac{8 \pi^2}{N},
\ee
on a volume $[-L/2,L/2]^3 \times [-T/2,T/2]$, with $T \gg L$. When
$T\rightarrow \infty$  these solutions represent vacuum to vacuum
tunneling.   

We have restricted our analysis to the following non-orthogonal
twist tensors:
\begin{enumerate}
\item Spatial twist, always $ \vec m = (1,1,1)$.
\item Temporal twist, two cases:
\begin{itemize}
\item $ \vec k = (1,0,0) $ for $N=3,4,5,8,10$ the solution is in 
this case anti-self-dual, $Q=-1/N$.
\item $ \vec k = (n,n,n) $ for $N=3n+1=4,7,10,13,19,25$ the solution 
is here self-dual, $Q=1/N$.   
\end{itemize}
\end{enumerate}

A list of all the lattices analyzed is presented in Table 1.

From the lattice configurations we can easily derive information concerning
continuum quantities. Part of  it can be extracted in a gauge invariant way,
such is the case for instance of the eigenvalues of the field strength
or the Polyakov loops. However  to derive information concerning
the gauge potential gauge fixing is needed.

The continuum field strength tensor is extracted, up to ${\cal O}(a^2)$,
from the clover average of the plaquette:
\be
 {\bf Q}_{\mu\nu}(n) =  \frac{1}{4} \clover
\ee
through
\be
{\bf F}_{\mu\nu}(na) = \frac{1}{a^2} \frac{1}{2i} \left[
{\bf Q}_{\mu\nu}(n)-{\bf Q}^{\dagger}_{
\mu\nu}(n) -
\frac{\Id}{\N}\Tr\left({\bf Q}_{\mu\nu}(n)-{\bf Q}^{\dagger}_{\mu\nu}(n)\right)
\right]
\ee
In terms of the gauge fixed links  the gauge potential is,
\be
{\bf A}_{\mu}\left[(n+\frac{1}{2})a\right] =
\frac{1}{a} \frac{1}{2i} \left[ {\bf U}^{gf}_{\mu}(n)
 -{\bf U}^{gf \dagger}_{\mu}(n) -
\frac{\Id}{\N}\Tr\left({\bf U}^{gf}_{\mu}(n)
 -{\bf U}^{gf \dagger}_{\mu}(n)\right) \right]  \label{eq:gp}.
\ee
The Polyakov loops ${\bf L}_{\mu}(x)$ are simply given  on the lattice by
the ordered product of the $\mu$-links corresponding to the path
$\gamma_{\mu}(x)$ in Eq.~(\ref{eq:pol}).

Non-gauge invariant information about the configurations will be 
presented in the
temporal gauge: ${\bf A}_4=0$. In addition we fix ${\bf A}_i(t=-\infty)=0$ which
is allowed because in the $T\rightarrow \infty$ limit fractional
instantons describe vacuum to vacuum tunneling.
In this gauge ${\bf A}_i(t=\infty)=-\imath {\bf \Omega}_4\partial_i
{\bf \Omega}_4^\dagger$, with ${\bf \Omega}_4$ the temporal twist matrix, and
the spatial twist matrices are constant.
 This is not yet a complete gauge fixing,
we still have the freedom to make a global gauge transformation
and also to multiply the twist matrices by an element of the center
of the group. We have made use of the global gauge transformation 
to bring the spatial twist matrices to a particular form. 
An explicit construction of constant spatial twist matrices compatible 
with the twist $\vec m = (1,1,1)$ can be easily found, following 't Hooft
\cite{thoof2}:
\be
{\bf \Omega}_3 = {\bf Q} \hspace{1.0 cm} {\bf \Omega}_2 = {\bf P}^{N-1} \hspace{1.0 cm}
{\bf \Omega}_1 = e^{\frac{i 2\pi p}{N}}\hspace{0.1 cm}{\bf PQ}^{N-1}  \label{mat}
\ee
with p an integer number taking values $p=1,2,...\N$, and ${\bf P}$, ${\bf Q}$ the
matrices,
\be
{\bf P} = \pmatrix{0&1&...&0\cr
               ...&...&...&...\cr 0&0&...&1\cr (-1)^{N+1}&0&...&0 }
\hspace{0.2cm}
{\bf Q} = e^{i\pi(1-N)/N}\pmatrix{\phi_0&0&...&0\cr 0&\phi_1&...&0\cr
                              ...&...&...&...\cr 0&0&...& \phi_{N-1}} \label{eq:pandq}
\ee
where $\phi_n= {\rm exp}(i2\pi n/N)$ with $n=0,1,...,\N-1$.

The invariance under multiplication by an element of the center of the
group is fixed by imposing that the Polyakov loops
take the value $A e^{i\pi}$
at the position where the energy density of the solution is maximal.

On the lattice the ${\bf A}_4=0$, ${\bf A}_i(t=-T/2)=0$ gauge is implemented by
transforming the corresponding link variables to the identity.
The gauge transformation, $\omega(n)$,
 which implements the change, is constructed in the following way:
choose a point in the time slice
$t=-T/2$, i.e. $n^0=(n_t=1,\vec{n}^0)$;
$\omega(n)$
is the product of the link variables along a certain path connecting
$n^0$ with
$n=(n_t,n_x,n_y,n_z)$. In particular we choose $n^0=(1,1,1,1)$
and the path such that it reaches the point $n$  first in the
$x$ direction up to $n_x$, then in the $y$ direction up to $n_y$, in
the
$z$ direction up to $n_z$ and finally in the $t$ direction up to $n_t$.
In this gauge the information about the twist matrices is encoded in
the links ${\bf U}_0(n_t=N_t)$,  ${\bf U}_i(n_t=1,n_i=N_s)$, the latter are 
rotated to the form indicated in Eq. (\ref{mat}) .

\subsection{Gauge-invariant quantities}
{\bf 1. Global quantities}. In Table 1 we give the values obtained for the action $S$,
electric and magnetic parts of the action, $S_e$ and $S_b$ respectively, 
and topological charge $Q$:
\bea
\lefteqn{S = \frac{1}{2} \int \Tr \left( {\bf F}_{\mu \nu} {\bf F}_{\mu \nu} \right) d_4x=
\int \Tr \left( {\bf E}^2_i + {\bf B}^2_i \right) d_4x } \hspace{3.5 in} \nonumber \\
\lefteqn{S_e = \int \Tr \left( {\bf E}^2_i\right) d_4x  \hspace{0.5 in}
S_b = \int \Tr \left( {\bf B}^2_i\right) d_4x  } \hspace{3.5 in} \nonumber \\
\lefteqn{Q = \frac{1}{16\pi^2} \int \Tr \left( {\bf F}_{\mu \nu} \tilde{{\bf F}}_{\mu \nu} \right) d_4x=
\frac{1}{4\pi^2}\int \Tr \left( \vec {\bf E} \vec {\bf B} \right) d_4x } \hspace{3.5 in} \label{eq:SSSQ}
\eea
where ${\bf E}_i={\bf F}_{4i}$ and $\half \epsilon_{ijk}{\bf F}_{ij}={\bf B}_k $. We can see from the data
that the configurations obtained are (anti-)self-dual to a very good degree, being 
therefore solutions of the Euclidean equations of motion. Those values are very 
near to the continuum expected values $S\N/8\pi^2=1$,
$S_e\N/8\pi^2=0.5$, $S_b\N/8\pi^2=0.5$ and $Q\N=\pm1$.

\linespread{1.0}
\begin{table}
\begin{center}
\vspace{-0.5 cm}
\caption{{\footnotesize Set of studied solutions. All quantities are defined on equation \ref{eq:SSSQ}. In the first column we label the solutions for reference in other tables.} }
\vspace{0.2 cm}
\begin{tabular}{||c|c|c|c||c|c|c|c||}
\hline
Sol. & Group  & $\vec k$ & Size  & $\frac{S \N}{8\pi^2}$ &  Q \N   & $ \frac{S_e \N}{8\pi^2}$  &  
$ \frac{S_b \N}{8\pi^2}$  \\   
\hline \hline
      & SU(3) & (1,0,0)   &  $6^3 \times 18 $  & 0.95409  & -0.95370  &  0.47817  &   0.47592   \\ \hline
      & SU(3) & (1,0,0)   &  $7^3 \times 21 $  & 0.96623  & -0.96602  &  0.48393  &   0.48230   \\ \hline
      & SU(3) & (1,0,0)   &  $8^3 \times 24 $  & 0.97413  & -0.97401  &  0.48768  &   0.48645   \\ \hline
  I.1 & SU(3) & (1,0,0)   & $10^3 \times 30 $  & 0.98343  & -0.98338  &  0.49210  &   0.49133   \\ \hline
  I.2 & SU(4) & (1,0,0)   &  $8^3 \times 32 $  & 0.98273  & -0.98267  &  0.49168  &   0.49105   \\ \hline
  I.3 & SU(5) & (1,0,0)   &  $8^3 \times 40 $  & 0.98742  & -0.98738  &  0.49387  &   0.49355   \\ \hline
  I.4 & SU(8) & (1,0,0)   &  $5^3 \times 40 $  & 0.98377  & -0.98368  &  0.49194  &   0.49183   \\ \hline
  I.5 &SU(10) & (1,0,0)   &  $6^3 \times 48 $  & 0.99204  & -0.99201  &  0.49465  &   0.49739   \\ \hline
 II.1 & SU(4) & (1,1,1)   &  $8^3 \times 32 $  & 0.98275  &  0.98268  &  0.49166  &   0.49109   \\ \hline
      & SU(7) & (2,2,2)   &  $3^3 \times 21 $  & 0.94440  &  0.94346  &  0.47284  &   0.47155   \\ \hline
      & SU(7) & (2,2,2)   &  $5^3 \times 35 $  & 0.98020  &  0.98008  &  0.49026  &   0.48994   \\ \hline
 II.2 & SU(7) & (2,2,2)   &  $8^3 \times 56 $  & 0.99229  &  0.99228  &  0.49613  &   0.49617   \\ \hline
 II.3 &SU(10) & (3,3,3)   &  $4^3 \times 40 $  & 0.98192  &  0.98180  &  0.49100  &   0.49092   \\ \hline
 II.4 &SU(13) & (4,4,4)   &  $4^3 \times 52 $  & 0.98801  &  0.98794  &  0.49399  &   0.49401   \\ \hline
      &SU(19) & (6,6,6)   &  $2^3 \times 38 $  & 0.97324  &  0.97200  &  0.48729  &   0.48594   \\ \hline
 II.5 &SU(19) & (6,6,6)   &  $3^3 \times 57 $  & 0.98812  &  0.98804  &  0.49405  &   0.49406   \\ \hline
 II.6 &SU(25) & (8,8,8)   &  $2^3 \times 50 $  & 0.98205  &  0.98187  &  0.49109  &   0.49096   \\ \hline
\end{tabular}
\end{center}
\label{tb:ressol}
\end{table}
\linespread{1.42}

The first thing we should check is the scaling of the solutions.  Without loss 
of generality we set the spatial length to $l_s=1$, being then the
lattice spacing $a=1/N_s$. To see how the continuum limit $a\rightarrow 0$ is 
approached we vary the lattice spacing while keeping all other parameters
fixed (among them the ratio $N_t/N_s$). We fit the $N$ and $a^2$ dependence
of the action to the expression  $S\N/8\pi^2= 1- \Delta a^2/(\N\sqrt{\N})$
and obtain that for the value $\Delta = 8.893$ the data in table 1 are well
described with errors smaller than the $0.2 \% $ . 
From this fit we understand how the continuum limit is approached 
for any value of $\N$, and also that the $\N$ dependence is such that  
the lattice corrections decrease with increasing $\N$. This property
will be discussed further later on, it implies that for large $\N$ we can obtain good 
continuum results already on rather coarse lattices, this is a rather
general property which, as we will see, affects other quantities apart 
from the integrated action and charge densities.

\vspace{.3cm}

{\bf 2. Energy profile}. The energy profile, defined as
\be
   \epsilon(t) = \int \Tr \left( E^2_i(\vec{x},t) + B^2_i(\vec{x},t) \right) 
d_3 \vec{x} , 
\ee
is located on a region of size  $\sim N/3$ and has only one maximum for all values of
N up to $N=13$ (instanton profile) and a double peak structure for the values $N=19,25$.
In Figures 1a and 1b we show the scaling with the lattice spacing for the solutions with 
$N=3$ and $N=19$. We can see that points coming from lattices with different sizes 
describe very similar curves, scaling towards the same continuum function.

For values of  N up to $N=13$ , $\epsilon(t)$ is well fitted by
\be
 \phi(t) = \frac{1}{A\hspace{0.1 cm}cosh(wt)+Bt^2+C}  \hspace{0.5 cm} . \label{fit}
\ee
The values obtained for the parameters $A,B,C,w$ are given in table 2.
For the values $N=19,25$ we fit to the expression $(\phi(t-t_0)+\phi(t+t_0))/2$ 
and also in table 2 we give the results of these fits.

\begin{figure}
\vbox{ \hbox{       \vbox{ \epsfxsize=3truein \hbox{\epsffile{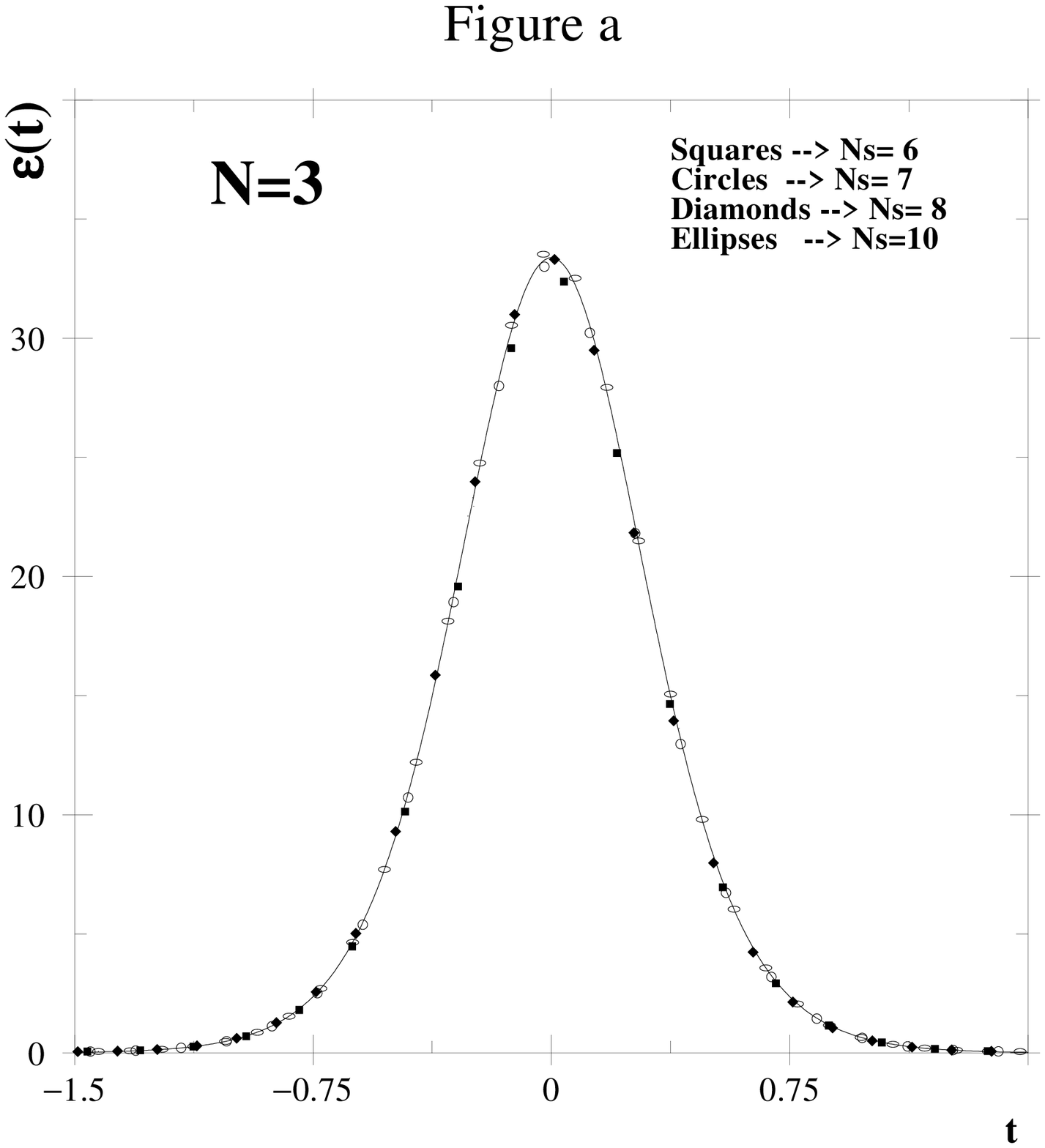} } }
             \hfill \vbox{ \epsfxsize=3truein \hbox{\epsffile{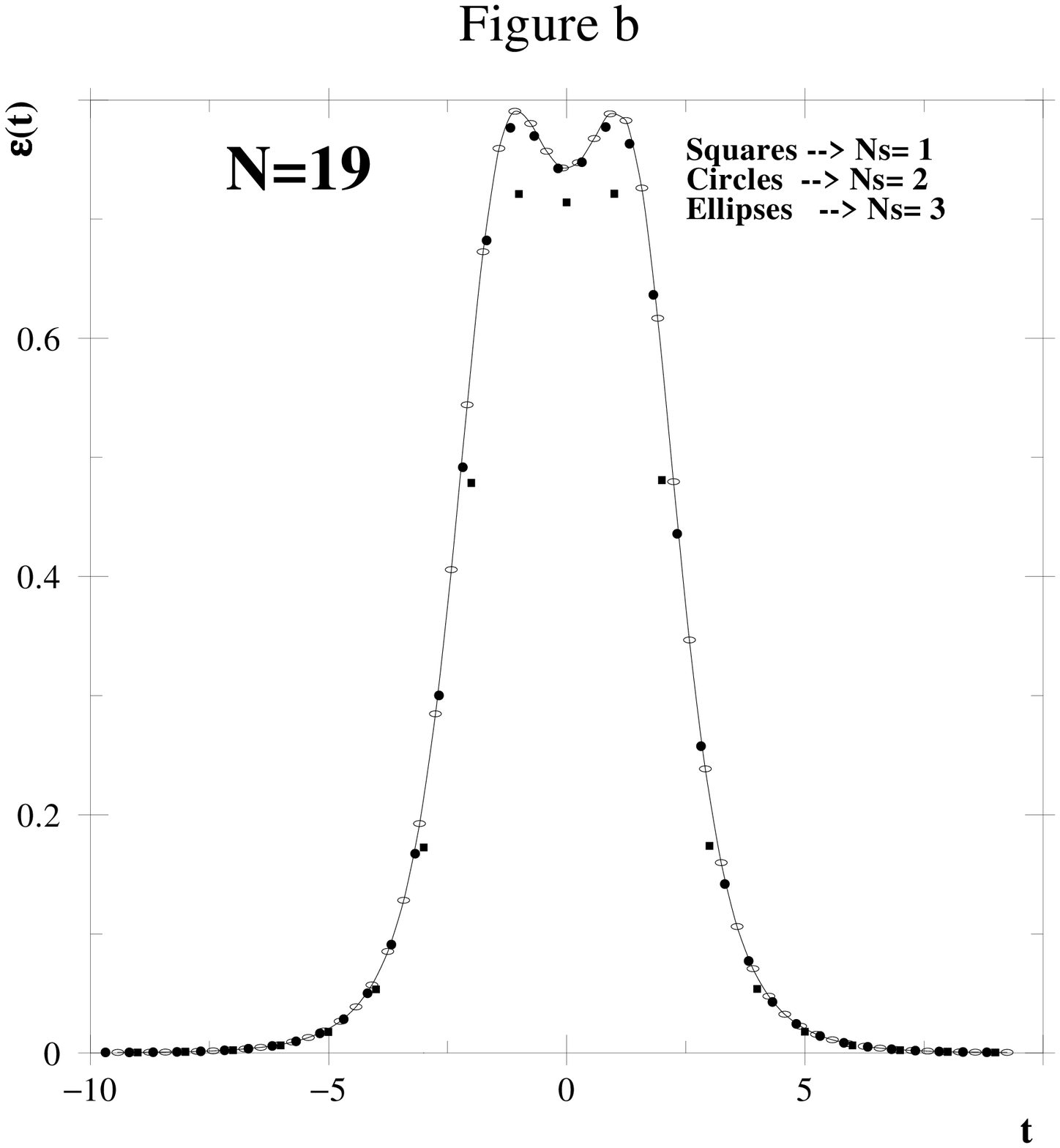} } }  }
       \hbox{       \vbox{ \epsfxsize=3truein \hbox{\epsffile{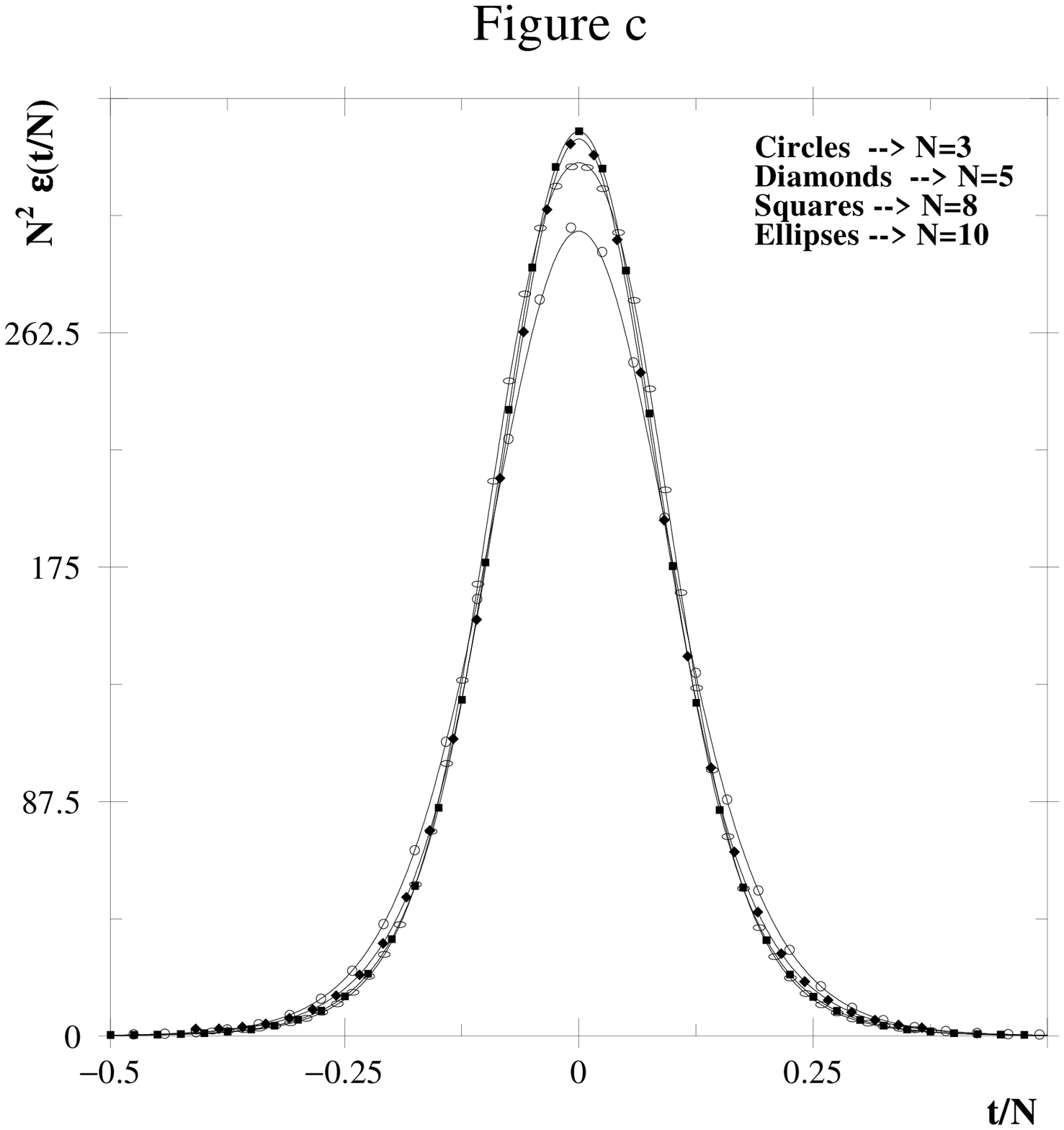}} }
             \hfill \vbox{ \epsfxsize=3truein \hbox{\epsffile{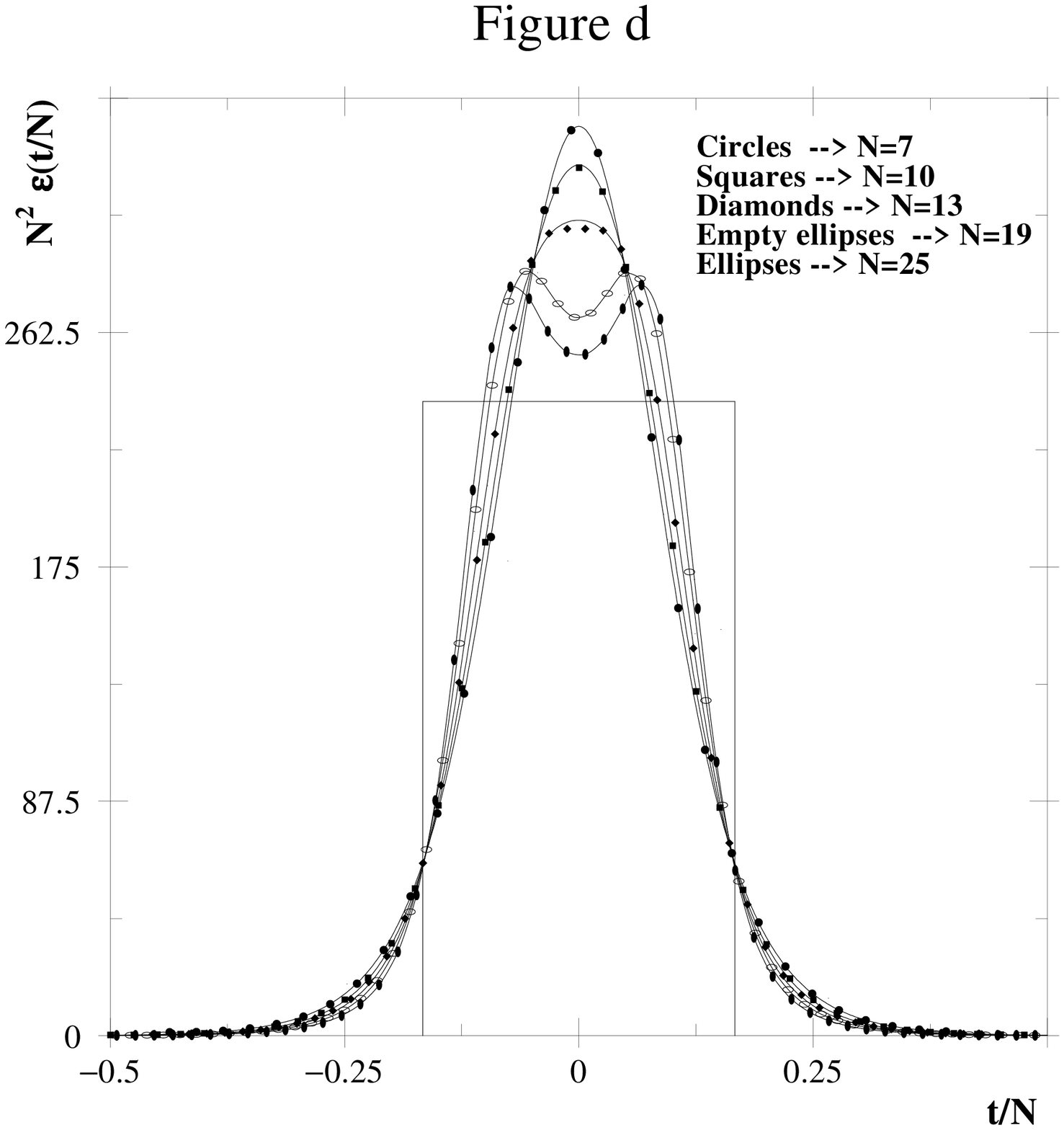}} }  }
     }
\caption{ {\footnotesize In figures a and b it is shown the energy profile as a
function of time calculated with different lattice sizes for the solutions with 
N=3 and N=19, respectively. 
In figures c and d it is shown the $N$ behaviour of the energy profile,
plotting $N^2$ times the value of the profile as a function of $t/N$, in figure
c for solutions with $\vec k =(1,0,0)$ and in figure d with $\vec k = (n,n,n)$. 
The continuum curves are the fits to the function \ref{fit} given in table 2.} }
\label{fig:perfil}
\end{figure}

The N dependence is such that
\be
   \epsilon(t) \simeq \frac{\phi(t/\N)}{\N^2}
\ee
as  illustrated in figures 1c and 1d where 
$\N^2\epsilon(t)$ is plotted as a function of $t/\N$ for
$\vec k = (1,0,0)$ and $\N=4,5,8,10$ in figure 1c and for
$\vec k = (n,n,n)$ and $\N=4,7,10,13,19,25$ in figure 1d. We also plot
in figure 1d the energy profile for the abelian solution described in the
appendix (equations \ref{eq:abseF},\ref{eq:abseA} and \ref{eq:abseO}). 
We plot $N^2 \epsilon(t) = 24\pi^2$ for the values of $t/N$ between
$t/N=-1/6$ and $t/N=1/6$. This is the profile for the selfdual abelian solution in the 
$\N \rightarrow \infty$ limit. We can see that the energy profiles of the solutions
with  $\vec k = (n,n,n)$ are approaching the one of the abelian solution in the 
$\N \rightarrow \infty$ limit.

\linespread{1.0}
\begin{table}
\begin{center}
\caption{ {\footnotesize Results of the fit to the energy profile.} }
\label{tb:ajperf}
\vspace{0.1 cm}
\begin{tabular}{||c|c||c|c|c|c|c|c||}
\hline
Sol. & N   &  A  &  B  &  C  & N w & $t_0$ & $\chi^2/N_t$ \\   
\hline \hline 
 I.1  & 3 & 0.00863 &  0.03009 & 0.02132 & 17.95  & 0.0  & 0.0001500 \\ \hline
 I.2  & 4 & 0.02371 & -0.01394 & 0.02588 & 17.10  & 0.0  & 0.0000468 \\ \hline 
 I.3  & 5 & 0.04700 & -0.04251 & 0.02766 & 16.51  & 0.0  & 0.0000490 \\ \hline 
 I.4  & 8 & 0.11399 & -0.07989 & 0.07554 & 17.54  & 0.0  & 0.0000016 \\ \hline 
 I.5  &10 & 0.22671 & -0.20144 & 0.08006 & 17.26  & 0.0  & 0.0000074 \\ \hline 
II.1  & 4 & 0.02357 & -0.01222 & 0.02602 & 17.11  & 0.0  & 0.0000454 \\ \hline
II.2  & 7 & 0.07820 & -0.03956 & 0.06456 & 17.63  & 0.0  & 0.0000037 \\ \hline
II.3  &10 & 0.17808 & -0.15184 & 0.12957 & 18.23  & 0.0  & 0.0000131 \\ \hline 
II.4  &13 & 0.41834 & -0.34723 & 0.13673 & 17.94  & 0.0  & 0.0000427 \\ \hline  
II.5  &19 & 1.79852 & -0.59304 &-1.08299 & 18.93  & 1.24 & 0.0000542 \\ \hline  
II.6  &25 & 2.02385 & -0.59827 &-0.82069 & 22.86  & 1.74 & 0.0000168 \\ \hline  
\end{tabular}
\end{center}
\end{table}
\linespread{1.42}

\vspace{.3cm}

{\bf 3. Action density}. Defined as,
\be
   S(\vec x,t) =  \Tr \left( {\bf E}^2_i(\vec{x},t) + {\bf B}^2_i(\vec{x},t) \right) \hspace{0.5 cm} . 
\ee
For values of N up to $N=13$, the action density has only one maximum what we will call 
the center of the instanton. 
We fit the center and their first nearest 
neighbours to the expression,
\be
   S(\vec x,t) =  S_0 \left( 1 - \sum_{i}\frac{(x_i-x_i^{0})^2}{a_i^2}
    -\frac{(t-t^0)^2}{a_t^2} \right)
\ee
where $S_0$ is the height, $x_i^{0}$,$t^0$ the position
and $a_i$,$a_t$ the width of the maximum.
The values obtained are shown in table 3. For the values $N=19,25$ we observe two
maximum in the action density and we make the same fit for each one. The results
are also shown in table 3. In both cases, the errors are obtained from the difference
between the same quantities calculated for the electric and magnetic part of the action.
We can see that with increasing N all maximum become spatially flat. In fact,
this is a general and important property of the solutions; when N is large 
some quantities, among them the action density, are
spatially independent. In particular this implies that all the coordinate dependence of
the action density comes through the time dependence of the
energy profile $\epsilon(t)$ defined above. This fact allows to easily
understand the decrease of the lattice artifacts with increasing N
since generally constant fields give rise to a much smoother,
continuum-like, behaviour. 

\linespread{1.0}
\begin{table}
\begin{center}
\caption{ {\footnotesize Sizes and heights of the maximum in the action density. } }
\vspace{0.1 cm}
\begin{tabular}{||c|c||c|c|c||}
\hline
Sol. & N  &  $ a_s $  &  $ a_t $ & $S_0$   \\   
\hline \hline 
 I.1 & 3  & 0.531(4)    &  0.389(2) &  63.0(2)    \\ \hline
 I.2 & 4  & 0.760(6)    &  0.484(5) &  27.56(12)  \\ \hline
 I.3 & 5  & 1.03(3)     &  0.583(4) &  15.68(12)  \\ \hline 
 I.4 & 8  & 2.90(4)     &  0.980(4) &  5.38(2)    \\ \hline   
 I.5 &10  & 6.0(2)      &  1.535(5) &  3.25(1)    \\ \hline 
II.1 & 4  & 0.763(10)   &  0.480(5) &  27.65(12)  \\ \hline
II.2 & 7  & 1.98(3)     &  0.814(3) &  7.29(4)    \\ \hline
II.3 &10  & 6.53(7)     &  1.51(1)  &  3.253(7)   \\ \hline 
II.4 &13  &19.5(1.5)    &  5.2(8)   &  1.7835(7)  \\ \hline 
II.5 &19  & 290(20)     &  2.3(2)   &  0.794(8) \\ \hline 
II.6 &25  &  -          &  2.4(2)   &  0.448(3)    \\ \hline  
\end{tabular}
\end{center}
\label{tb:anch}
\end{table}
\linespread{1.42}

\vspace{.3cm}
{\bf 4. Eigenvalues of ${\bf F}_{\mu\nu}$}.

Since the solution is (anti) self-dual we only give the results for ${\bf B}_i$. 
The main properties for the eigenvalues of ${\bf F}_{\mu\nu}$ are illustrated in 
figures 2a, 2b, 2c and 2d. We only show the results for the solutions with twist
$\vec{k}=(n,n,n)$ because the same properties are obtained for the solutions 
with twist $\vec{k}=(1,0,0)$.

      In figure 2a we show the eigenvalues of ${\bf B}_1$ for the solutions with $N=7$ 
and $\vec k = (2,2,2)$. Very similar results are obtained if we plot ${\bf B}_2$ or ${\bf B}_3$
instead of ${\bf B}_1$. That we plot is the spatial average of each eigenvalue as a function
of time. The error bars mean spatial dispersion of the eigenvalues (difference
between the maximum and minimum value of the eigenvalue at each temporal point).
The first property we observe is spatial independence of the eigenvalues. This
property also holds for values of $N \geq 7$ as shown in figures 2b, 2c and 2d. 
From figure 2a we observe that we obtain good results from very coarse lattices. 
In this figure we plot points coming from lattices with the following sizes: 
$3^3 \times 21 $, $5^3 \times 35$ and $8^3 \times 56$, being the results almost
independent of the lattice size. This property also holds for bigger values of $N$.

\begin{figure}
\vbox{ \hbox{       \vbox{ \epsfxsize=3truein \hbox{\epsffile{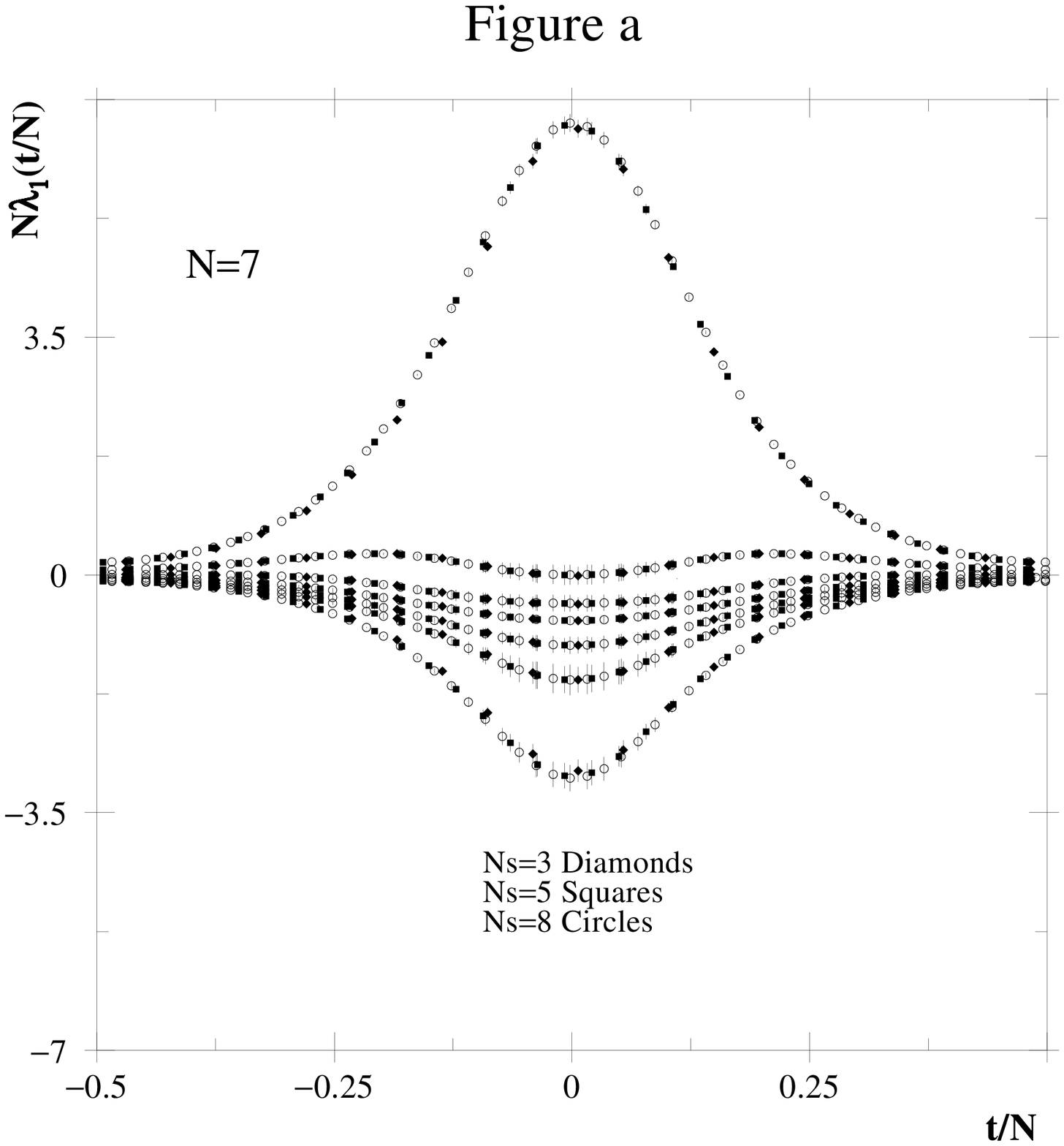} } }
             \hfill \vbox{ \epsfxsize=3truein \hbox{\epsffile{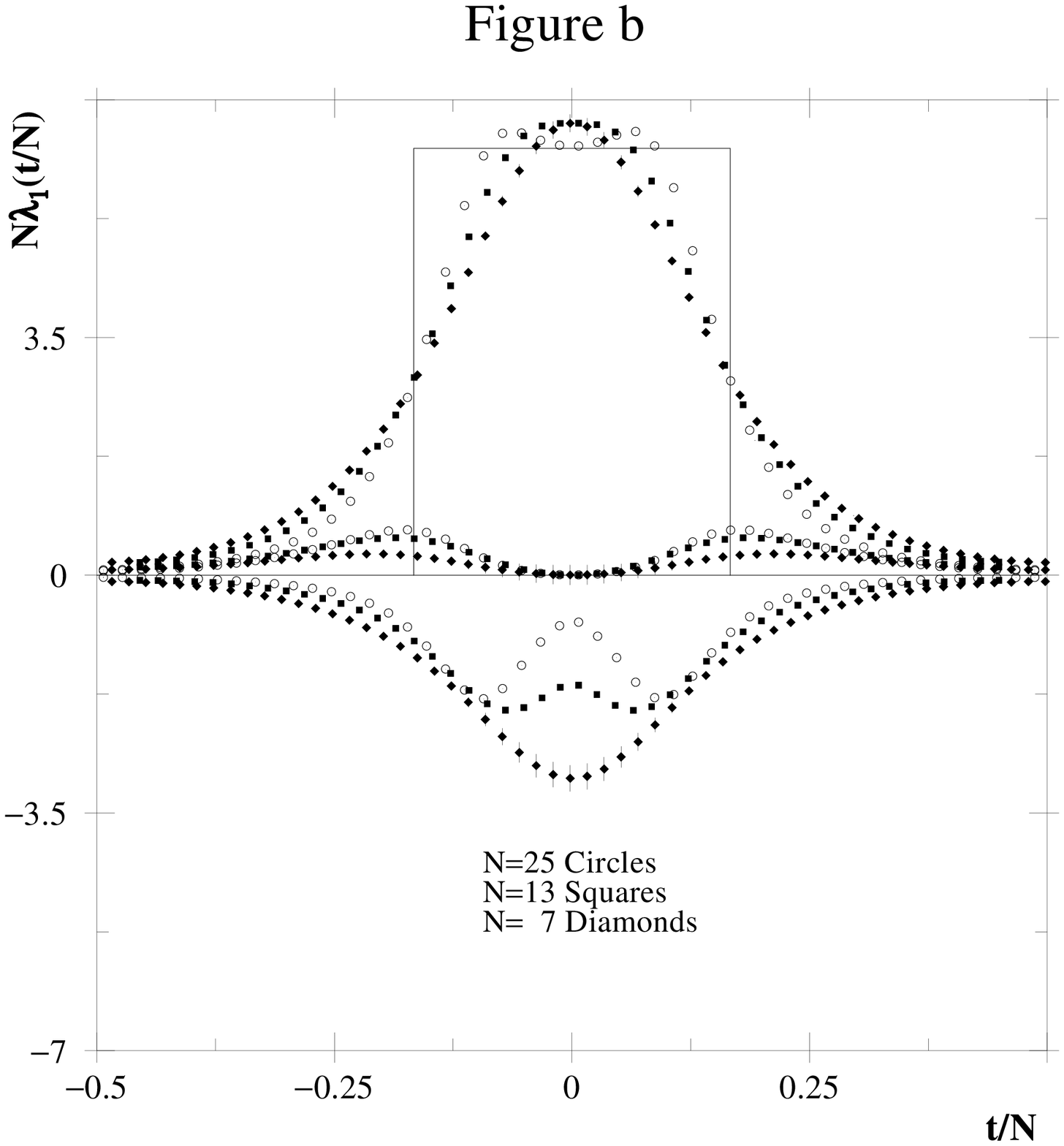} } }  }
       \hbox{       \vbox{ \epsfxsize=3truein \hbox{\epsffile{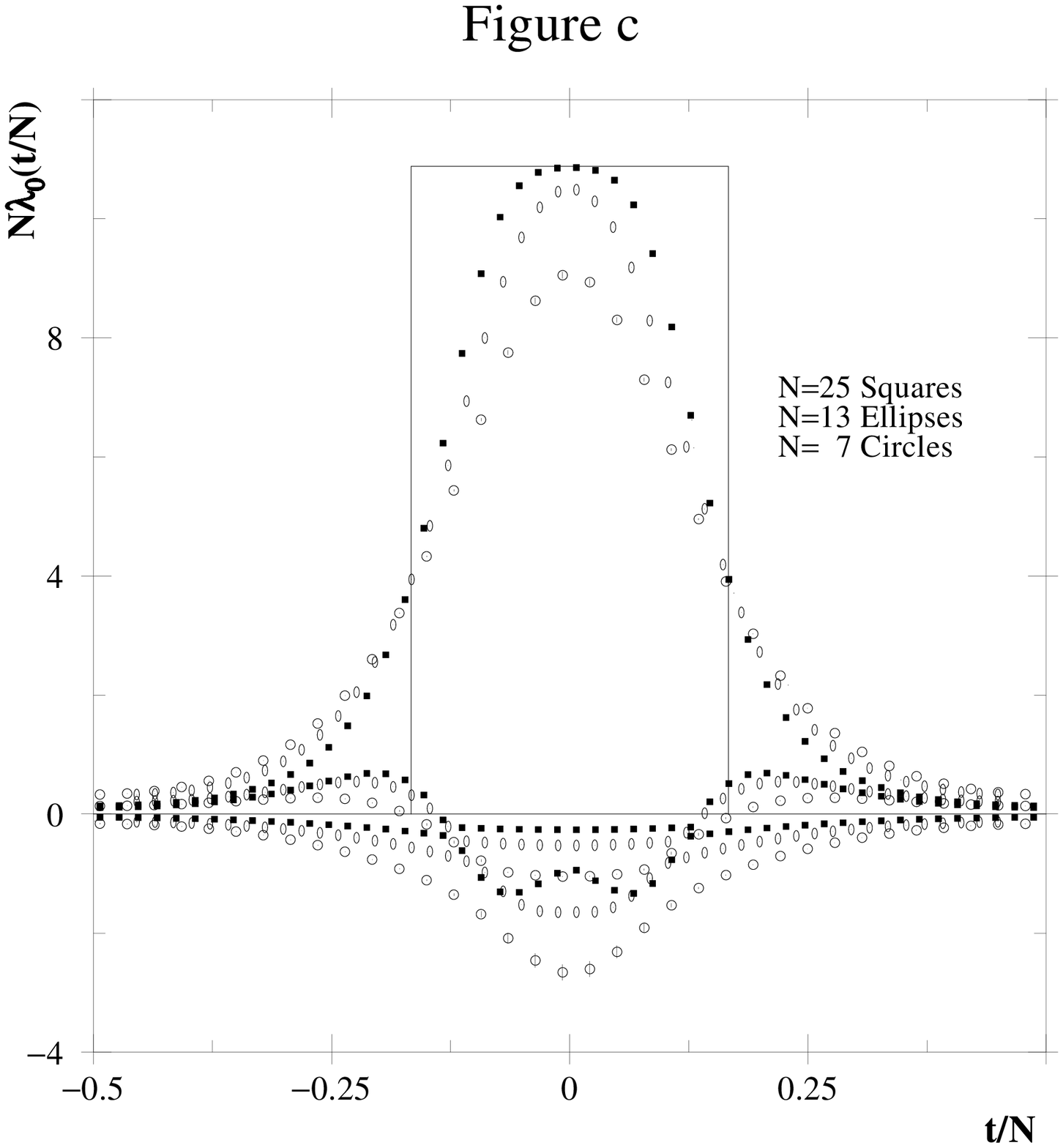}} }
             \hfill \vbox{ \epsfxsize=3truein \hbox{\epsffile{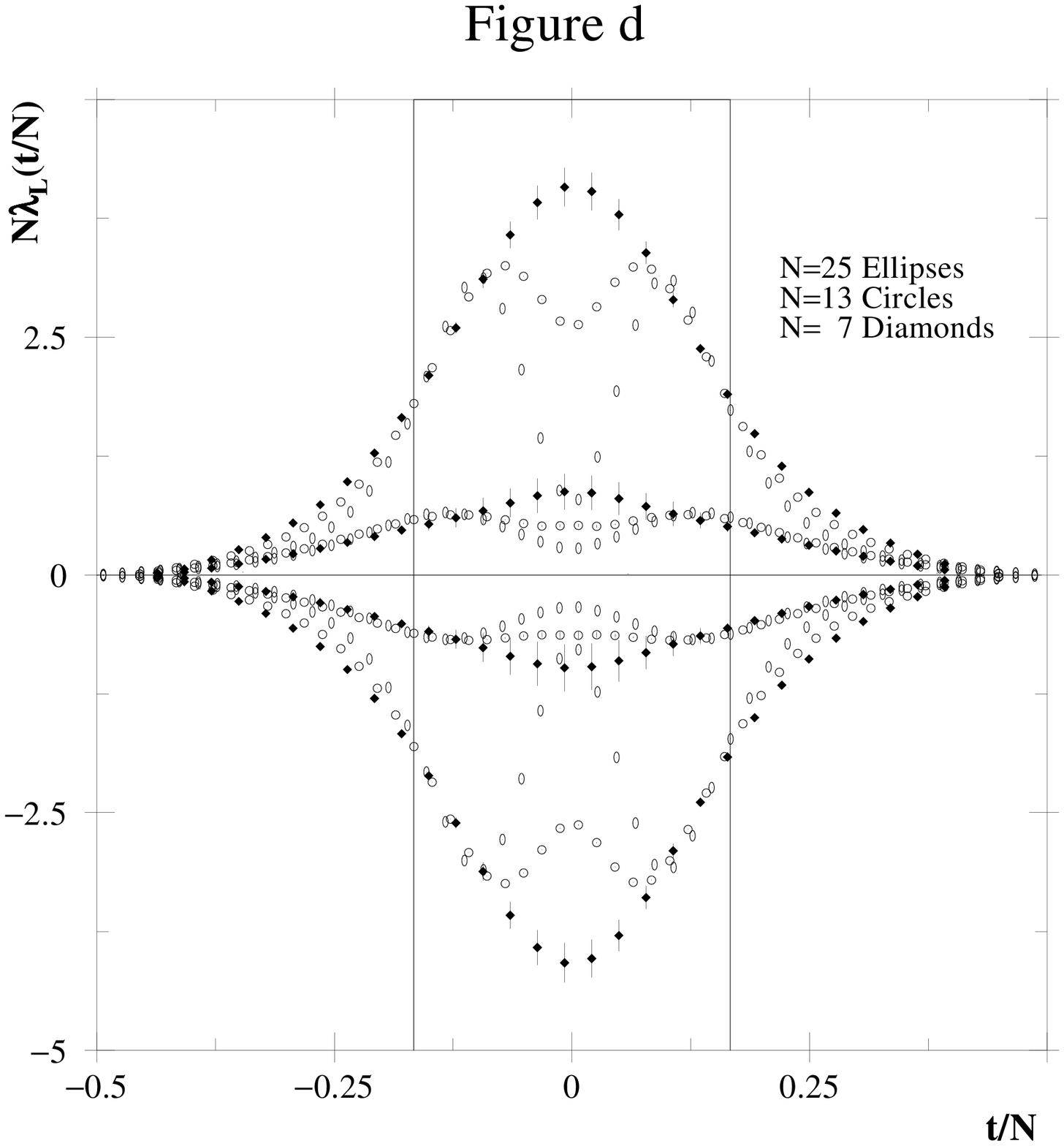}} }  }
     }
\caption{ {\footnotesize In figure a it is shown the eigenvalues $\lambda_1$ 
of ${\bf B}_1$ as a function of time, for the solution with $N=7$ 
and different lattice sizes.  To compare with the other figures $\lambda_1$ 
is multiplied by $N$ and $t$ divided by $\N$.  
In figure b  we plot the eigenvalues $1,2$ and $N$ of ${\bf B}_1$ multiplied
by N as a function of $t/N$, for three values of $N$. In figure c 
the same as in figure b, in this case for ${\bf B}_0$. In figure d it is shown 
the eigenvalues $1,2,N-1$ and $N$ of ${\bf B}_L$ for the same solutions as
in figure b.} }
\label{fig:fmunu}
\end{figure}

The behavior with N is shown in Figure 2b in which we plot the first, second
and $\N^{th}$ eigenvalues of $\N \times {\bf B}_1$ as a function of $t/\N$ 
for $\N=7,13,25$ and $\vec k = (n,n,n)$. The meaning of the points and the
error bars is the same as in figure 2a. The first eigenvalue is approximately
independent of N and the other N-1 become degenerate for increasing N. This structure
is very similar to the one of the selfdual abelian solution described in
the appendix (equations \ref{eq:abseF}, \ref{eq:abseA} and \ref{eq:abseO})  , 
the first eigenvalue of $N \times B_1$ takes the 
value $N\times \frac{2 \pi}{N}$ and the other $N-1$ are equal to the value 
$- N \times \frac{2 \pi}{(N-1)N}$. To compare we plot on figure 2b the first
eigenvalue for the abelian solution in the $N \rightarrow \infty $ limit, the
value $2\pi$ for the interval $-1/6 \leq t/N \leq 1/6$. We see that  
the first eigenvalue of $B_i$ for the solutions with  $\vec k = (n,n,n)$ 
are approaching the one of the abelian solution in the 
$\N \rightarrow \infty$ limit.

The symmetry properties of the spatial twist vector $\vec m$ suggest us 
to consider the following combinations of $B_i$ fields,
\bea
\lefteqn{ {\bf B}_0 = \frac{1}{\sqrt{3}} \left(   {\bf B}_1 + {\bf B}_2 + {\bf B}_3  \right)     } \hspace{5.0 cm} \nonumber \\ 
\lefteqn{ {\bf B}_L = \frac{1}{\sqrt{6}} \left( -2{\bf B}_1 + {\bf B}_2 + {\bf B}_3  \right)     } \hspace{5.0 cm} \nonumber \\ 
\lefteqn{ {\bf B}_T = \frac{1}{\sqrt{2}} \left(       - {\bf B}_2 + {\bf B}_3  \right)     } \hspace{5.0 cm}   
\label{eq:compabel}
\eea
one parallel to the $\vec m$ vector and the other two perpendicular. Also interesting is
that, if there is a common component in color space for $B_i$ fields, we will see this component
appearing in $B_0$ and not in $B_L$ and $B_T$.

In figure 2c we show the first, second and $\N^{th}$ eigenvalues of $N \times {\bf B}_0$ 
as a function of $t/N$ for the same solutions appearing in figure 2b (with the same
meaning for points and error bars). We observe the same structure seen in ${\bf B}_i$
and the expected property if there is one common component in color space,
the dominant eigenvalue is bigger than the one for ${\bf B}_i$. We can also observe that
this eigenvalue of $N \times {\bf B}_0$ is approaching to the shape of the one for the abelian
selfdual solution in the $N \rightarrow \infty$ limit, in this case $2 \pi \sqrt{3}$.

In figure 2d  we show the first, second, $(N \!\! - \! 1)^{th}$ and $N^{th}$ eigenvalues of
$N \times {\bf B}_L$ as a function of $t/N$ for the same solutions of figure 2b (also with
the same meaning for points and error bars). The same results are obtained if we show
${\bf B}_T$ instead of ${\bf B}_L$. The eigenvalue structure is completely different to the ones 
shown before. In this case the eigenvalues are distributed in pairs, each pair with
two opposite values. Another interesting property is that at $t/N=0$ the eigenvalues go
to zero very fast for large $N$, being ${\bf B}_0$ the only one non trivial in this limit.

\vspace{.3cm}

\begin{figure}
\caption{ { \footnotesize In figure a the eigenvalues ${\lambda}_{12}$ 
of $i[{\bf B}_1,{\bf B}_2]$ multiplied by $N^{3/2}$ are plotted as a function of $t/N$ for
three values of $N$. In figure b the quantity defined in equation 
\ref{eq:cosa} $cos^2 ( \alpha_{ij} )$ for $i=1$, $j=2$ is plotted as a function of $t/N$
for the solutions with $\vec k = (n,n,n)$. } }
\vbox{ \hbox{       \vbox{ \epsfxsize=3truein \hbox{\epsffile{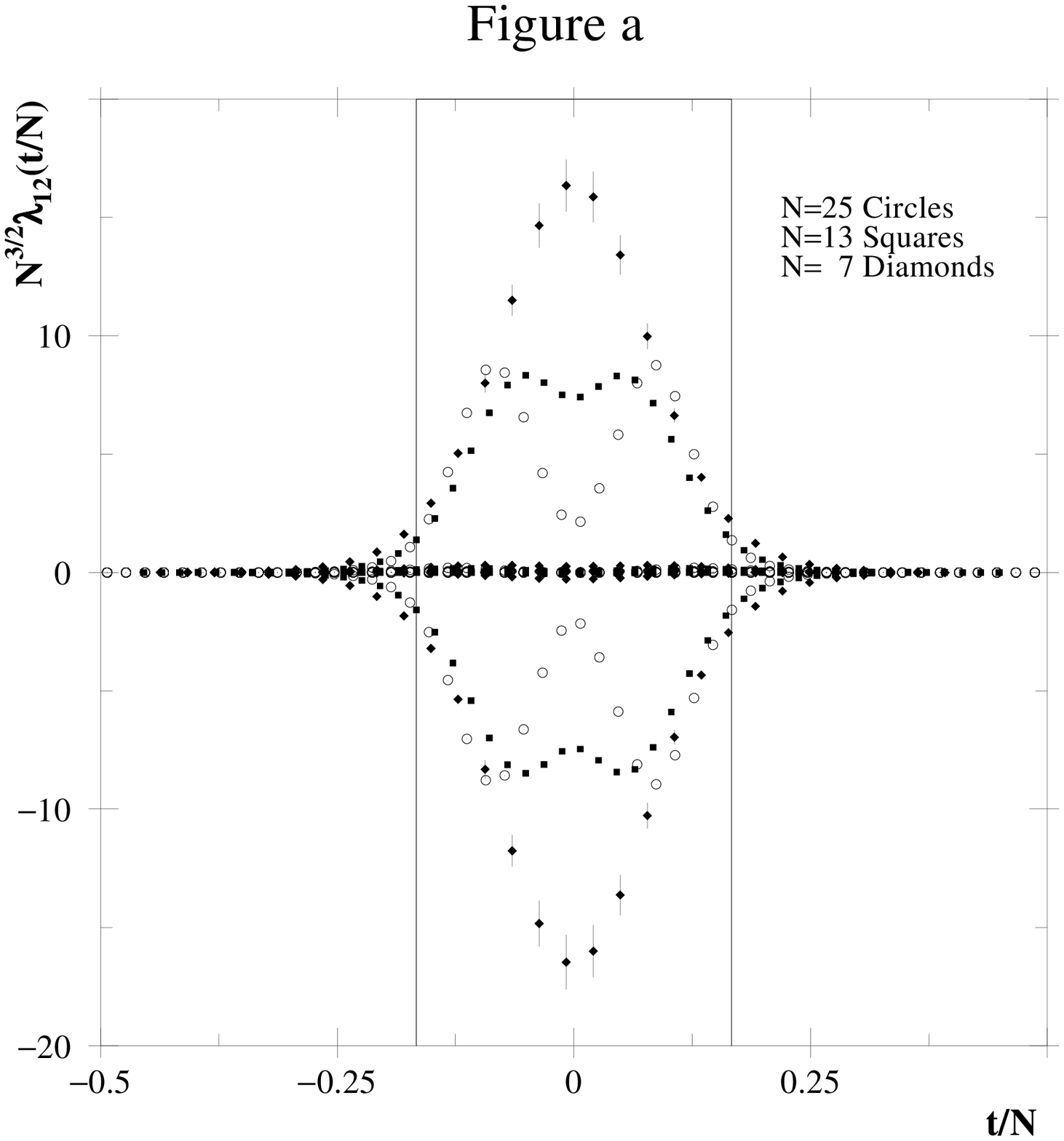} } }
             \hfill \vbox{ \epsfxsize=3truein \hbox{\epsffile{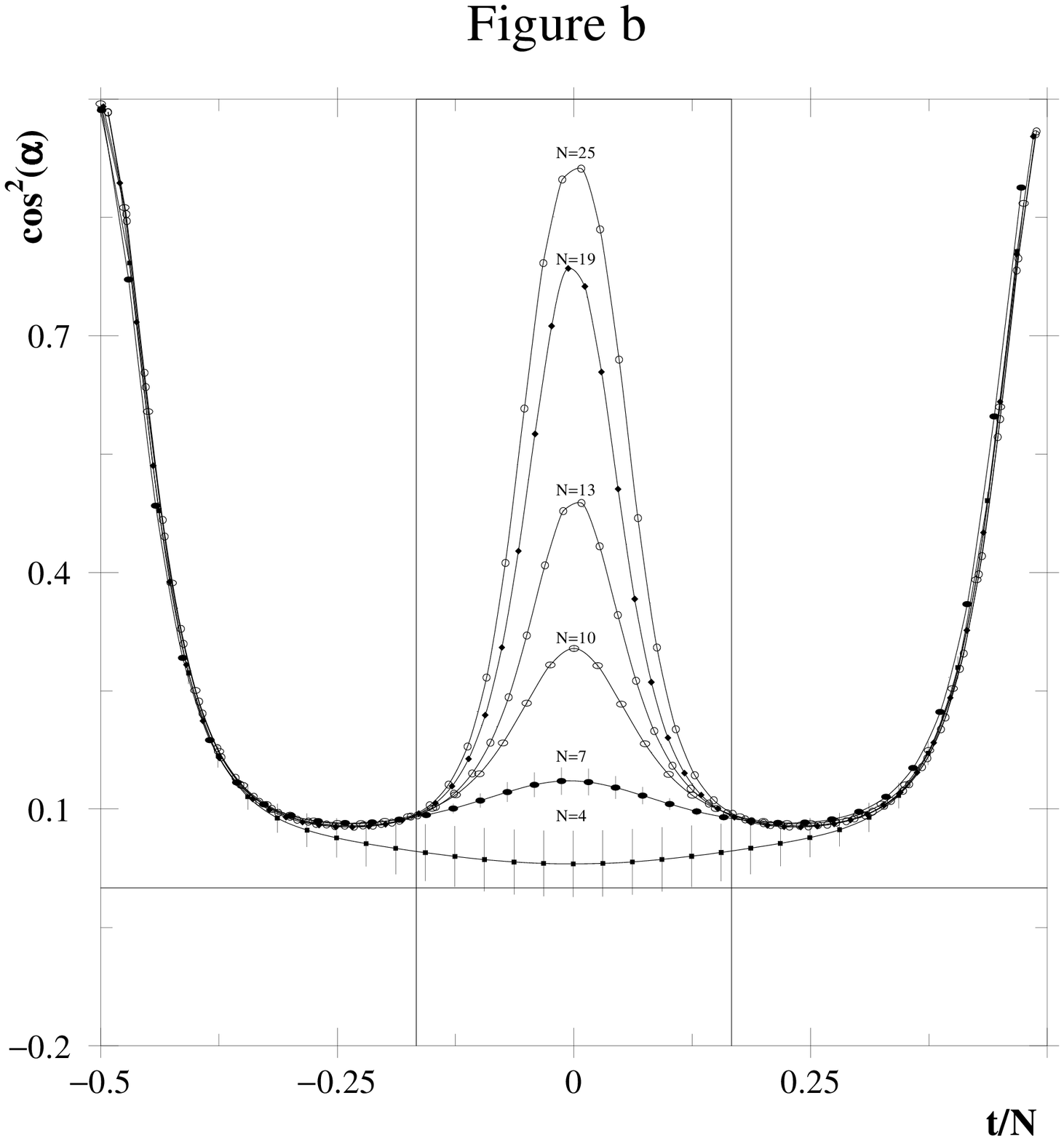 } } }
            }
     }
\label{fig:noab}
\end{figure}

{\bf 5. Colour orientation of the field strength}. This is studied by
calculating the commutators of $ {\bf B}_i$'s fields.
We show in Figure 3a the eigenvalues of $N^{3/2}[{\bf B}_1,{\bf B}_2]$ 
as a function of $t/\N$ for the solutions of Figure 2b,c,d (again the error
bars represent spatial dispersion). The same results are obtained if we plot
the eigenvalues of $N^{3/2}[{\bf B}_2,{\bf B}_3]$ or $N^{3/2}[{\bf B}_1,{\bf B}_3]$
instead of $N^{3/2}[{\bf B}_1,{\bf B}_2]$.
Only two eigenvalues are relevant being the other N-2 very close to zero.
We can see a nice scaling with $N^{3/2}$ of the two relevant eigenvalues at
points with $|t|/N>0.1$, but not at the center of the solution in which
the approach to zero is faster.  
For these quantities the spatial independence for large N  also holds.

To study the abelian content of ${\bf F}_{\mu\nu}$ we calculate the quantity:
\be
cos^2 \left( \alpha_{ij} \right) = \frac{tr({\bf B}_i {\bf B}_j)^2}{tr( {\bf B}_i)^2 tr({\bf B}_j)^2}  
\label{eq:cosa}
\ee
Whenever $cos^2 \left( \alpha_{ij} \right)=1$ the solution is abelian.
In Figure 3b we plot $cos^2 \left( \alpha_{12} \right)$ as a function
of $t/\N$ for the solutions with $\N=4,7,10,13,19,25$ and $\vec k = (n,n,n)$.
As in other figures, the points mean the spatial average of the quantity and
errors the spatial dispersion. For large N the solutions become
abelian at the instanton center (t=0). $cos^2 \left( \alpha_{ij} \right)$ shows again $\vec{x}$
independence for large N.

For the solutions with $\vec k = (1,0,0)$ we obtain the same behaviour near
the instanton center, but for $T\rightarrow \pm \infty$, $cos^2 \left( \alpha_{ij} \right)$
goes towards a $\N$ depending constant.

{\bf 6. Polyakov Loops and structure of vacuum}. 
 In the gauge we have chosen the relationship between the Polyakov loops 
and the twist matrices is specially clear.  We have ${\bf A}_4\!=\!0$
and hence the temporal Polyakov loop directly provides the twist matrix
${\bf \Omega}_4(\vec{x})$. At $t\!=\!-\infty$  ${\bf A}_i$  is also 
fixed to zero and in consequence
\be
 {\bf L}_i(\vec{x}, t=-\infty) =  {\bf \Omega}_i
\ee
Since the spatial twist matrices are constant, compatibility with the spatial
boundary conditions for $\vec{m}\!=\!(1,1,1)$ (see Eq.~(\ref{eq:bcpl})),
implies that $\N{\cal L}_{\mu}(t\!=\!-\infty)\!=\!\Tr( {\bf L}_{\mu}(t\!=\!-\infty))\!=\!0$.
From the boundary conditions in the time direction: 
\be
{\cal L}_i(\vec{x},t=T/2) = {\rm exp} \left(i k_i \frac{2\pi}{\N} 
\right) \hspace{0.2 cm} {\cal L}_i(\vec{x},t=-T/2) \quad ,\nonumber 
\ee
it is clear that also at $t\!=\!\infty$ the spatial Polyakov loops are zero.
To characterize the vacuum states between which the configurations interpolate we 
need thus an additional quantity provided by $\N {\cal L}_{zyx}=
\Tr({\bf L}_z {\bf L}_y {\bf L}_x)$. Using the twist matrices given in Eq.~(\ref{mat})
the values for  ${\cal L}_{zyx}$ in a vacuum are,
\be
{\cal L}_{zyx} = {\rm exp}(\imath 2\pi p/\N)
\ee 
where $p$ takes the values $p=1,...,\N$. There are therefore 
N different vacua labeled by the value of ${\cal L}_{zyx}$. Our solutions interpolate 
between two of them as can be seen from the boundary condition in the time 
direction for ${\cal L}_{zyx}$:
\be
{\cal L}_{zyx}(x_j,t=T/2) = {\rm exp} \left( \imath \sum_i k_i \frac{2\pi}{\N}
\right)
 \hspace{0.2 cm} {\cal L}_{zyx}(x_j,t=-T/2) .
\ee

We can parametrize the data obtained for ${\cal L}_i$, ${\cal L}_0$, ${\cal L}_{zyx}$ as,
\bea
\lefteqn{ {\cal L}_i(x_{j\neq i},t) = f_i(t) \hspace{0.2 cm} e^{i \alpha_i (t) }
\hspace{0.2 cm} \exp \left\{ \frac{i 2 \pi}{\N}
 (\vec m \times \vec r)_i \right\} } \hspace{7.0 cm} \label{eq:polaj1}    \\
\lefteqn{ {\cal L}_0(\vec{x}) = A \hspace{0.2 cm} e^{i \alpha_0 } \hspace{0.2 cm} \exp \left\{ \frac{-i 2 \pi}{\N}
 \vec k \vec r \right\} } \hspace{7.0 cm}  \label{eq:polaj2} \\
\lefteqn{{\cal L}_{zyx}(\vec{x},t) = f_{zyx}(t) \hspace{0.2 cm} e^{i \alpha_{zyx}(t) }} \hspace{7.0 cm} 
\label{eq:polaj3}
\eea

\noindent
note that with the gauge fixing condition for the Polyakov loops
those functions take the values: $\alpha_0\!=\!\pi$, $\alpha_i(t\!=\!0)=\pi$
($\vec r\! =\!\vec 0$ is the maximum of the solution). 

\linespread{1.0}
\begin{table}
\begin{center}
\caption{ {\footnotesize  Results of the fits to equations \ref{eq:polaj2}, \ref{eq:poli} 
and \ref{eq:polxyz}. } }
\vspace{0.1 cm}
\begin{tabular}{||c|c||c|c||c|c|c||c|c|c|c||}
\hline
Sol. & N  & $A$ & $\frac{10^4\chi^2}{N_s^3}$ & $B$  & $w$ & $\frac{10^3\chi^2}{N_t}$   & $ A_{zyx} $ & $ w_{zyx} $ & 
$ v_{zyx} $ & $\frac{10^4\chi^2}{N_s^3 \times N_t}$ \\   
\hline \hline 
 I.1 & 3  & 0.453 & 334.1        & 0.657  & 8.04 & 0.0188 & 2.057  & 15.77 & 20.52 & 13.898 \\ \hline
 I.2 & 4  & 0.472 & 75.41        & 0.764  & 8.20 & 0.0125 & 2.093  & 15.79 & 16.87 & 0.5014 \\ \hline
 I.3 & 5  & 0.485 & 12.96        & 0.846  & 8.22 & 0.0077 & 2.099  & 15.64 & 15.70 & 0.4166 \\ \hline 
 I.4 & 8  & 0.504 & 0.4252       & 0.969  & 8.35 & 0.0221 & 2.095  & 15.93 & 14.88 & 0.0411 \\ \hline 
 I.5 &10  & 0.508 & 0.0094       & 0.998  & 7.98 & 0.0214 & 2.092  & 16.02 & 14.60 & 0.0375 \\ \hline 
II.1 & 4  & 0.288 & 0.2311       & 0.763  & 8.11 & 0.0534 & 2.094  & 15.82 & 16.88 & 0.4846 \\ \hline
II.2 & 7  & 0.150 & 0.0017       & 0.929  & 8.11 & 0.0025 & 2.099  & 15.94 & 15.07 & 0.0868 \\ \hline
II.3 &10  & 0.102 & $< 10^{-6}$  & 0.999  & 8.19 & 0.0334 & 2.093  & 16.19 & 14.70 & 0.0353 \\ \hline 
II.4 &13  & 0.078 & $< 10^{-6}$  & 1.029  & 8.05 & 0.0925 & 2.095  & 16.53 & 14.52 & 0.0428 \\ \hline 
II.5 &19  & 0.053 & $< 10^{-6}$  & 1.057  & 7.96 & 0.2738 & 2.103  & 17.07 & 14.32 & 0.0425 \\ \hline 
II.6 &25  & 0.040 & $< 10^{-6}$  & 1.066  & 7.88 & 0.4323 & 2.123  & 17.45 & 14.21 & 0.0389 \\ \hline 
\end{tabular}
\end{center}
\label{tb:reloda}
\end{table}
\linespread{1.42}

For ${\cal L}_i$ we make a fit at each temporal point to the spatial dependence 
of equation \ref{eq:polaj1}. The values of $\sqrt{\chi^2/{N_s}^3}$
obtained for the solutions with temporal twist vector $\vec{k} = (n,n,n)$ are always smaller 
than $1.24^o$, $0.27^o$, $0.07^o$ and $0.004^o$ for the values $N=4,7,10$ and $13$ 
respectively. For the solutions with temporal twist vector $\vec{k}=(1,0,0)$ these
values are $2.7^o$, $1.28^o$, $0.81^o$, $0.13^o$ and $0.05^o$ for $N=3,4,5,8$ and
$10$ respectively. We extract the values of ${\cal L}_i$ at the spatial
maximum of the solution and make a fit to the expression:
\bea
\lefteqn{ f_i(t) = \frac{B}{N}cosh(w \hspace{0.1 cm} t/N) } \hspace{7.0 cm}  \label{eq:poli}
\eea
the values of $B$ and $w$ are given in table 4. 

The functions $f_{zyx}$, $\alpha_{zyx}^{(1,0,0)}$ and $\alpha_{zyx}^{(n,n,n)}$
parametrizing  ${\cal L}_{zyx}$ in equation $\ref{eq:polaj3}$ are well fitted by: 
\bea
\lefteqn{ f_{zyx}(t) = 1-\frac{A_{zyx}}{N \hspace{0.1 cm} 
          cosh(w_{zyx}\hspace{0.1 cm}t/N) } } \hspace{7.0 cm}\nonumber \\ 
\lefteqn{ \alpha_{zyx}^{(1,0,0)}(t) = \frac{\pi}{N}\left(1+tagh(v_{zyx}
          \hspace{0.1 cm}\frac{t}{N})\right) } \hspace{7.0 cm} \nonumber \\ 
\lefteqn{ \alpha_{zyx}^{(n,n,n)}(t) = \frac{\pi}{N}\left(1-tagh(v_{zyx}
          \hspace{0.1 cm}\frac{t}{N})\right)  } \hspace{7.0 cm} \label{eq:polxyz}  
\eea
\noindent
the values obtained for $A_{zyx}$, $w_{zyx}$, $v_{zyx}$ are given in table 4. And
finally ${\cal L}_{0}$ only needs the constant $A$ to be fitted, their values are given
in table 4.

We compare our results with the Polyakov loops for the abelian selfdual solution
described in the appendix,
\bea
\lefteqn{ {\cal L}_{i} (x_{j\neq i},t) =
                 \frac{1}{N} \hspace{0.2 cm} e^{i \pi }\hspace{0.2 cm}
                  Exp \left(  i \frac{2 \pi}{N} \frac{ N-1}{3} \frac{t}{T}\right)
                  \hspace{0.2 cm} Exp \left(  i \frac{2 \pi}{N} (\vec m \times \vec x)_i \right) }
                             \hspace{9.0 cm}\nonumber \\
\lefteqn{ {\cal L}_0(\vec{x}) = \frac{1}{N}
                  \hspace{0.2 cm} e^{i \pi } \hspace{0.2 cm} Exp \left\{ \frac{-i 2 \pi}{\N}
\vec k \vec x \right\} } \hspace{9.0 cm} \nonumber 
\eea 
where $-T/2 \leq t \leq T/2$ and $-0.5 \leq x_i \leq 0.5$.  We can see that the Polyakov loops
in the $N \rightarrow \infty$ limit for the solutions with $\vec k = (n,n,n)$ are the same as 
the ones for the abelian selfdual solution.

\subsection{Gauge-dependent quantities}

{\bf 1. Eigenvalues of ${\bf A}_i$}. After gauge fixing to the gauge described
previously, we calculate the eigenvalues of ${\bf A}_i$. The main properties of
these quantities are shown in figures 4a, 4b, 4c and 4d. We only show the results 
for the solutions with twist $\vec{k}=(n,n,n)$, because the same figures are 
obtained for the solutions with twist $\vec{k}=(1,0,0)$ changing the sign of the
gauge field, ${\bf A}_i \rightarrow -{\bf A}_i $.

In figure 4a we show how the eigenvalues of ${\bf A}_1$ scale towards the 
continuum limit. We plot these quantities for the solutions with $N=7$, temporal
twist vector $\vec{k} = (2,2,2)$ and lattice sizes $3^3 \times 21$, $5^3 \times 35$
and $8^3 \times 56$. As in previous figures, points values are the spatial average of
the eigenvalues and error bars mean spatial dispersion. 
We can see that the discretization errors still are important for
the solution with size $N_s = 3$ but are very small for bigger sizes. For
these quantities we consider in the following lattice sizes with $N_s \geq 4$.

\begin{figure}
\vbox{ \hbox{       \vbox{ \epsfxsize=3truein \hbox{\epsffile{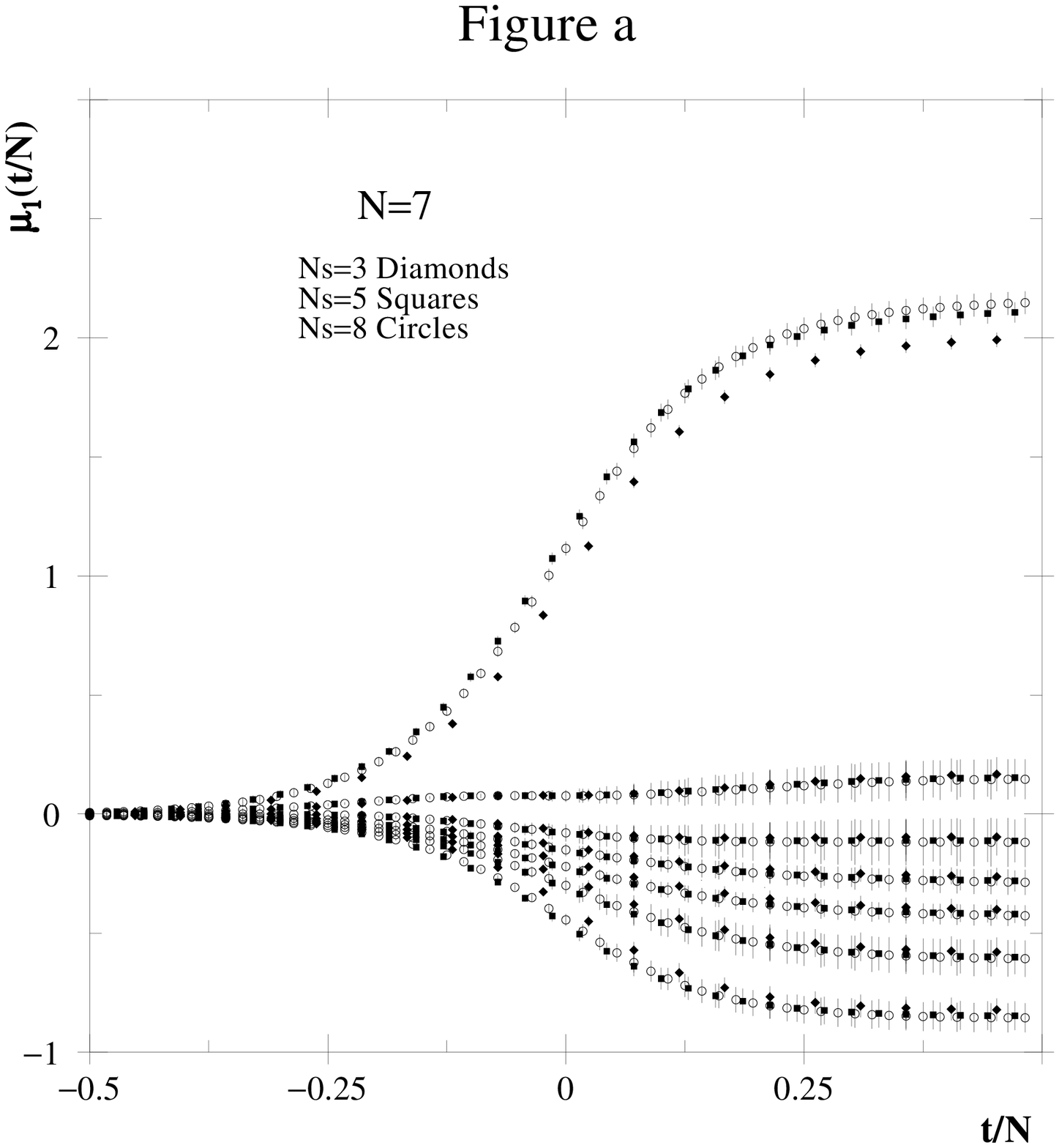} } }
             \hfill \vbox{ \epsfxsize=3truein \hbox{\epsffile{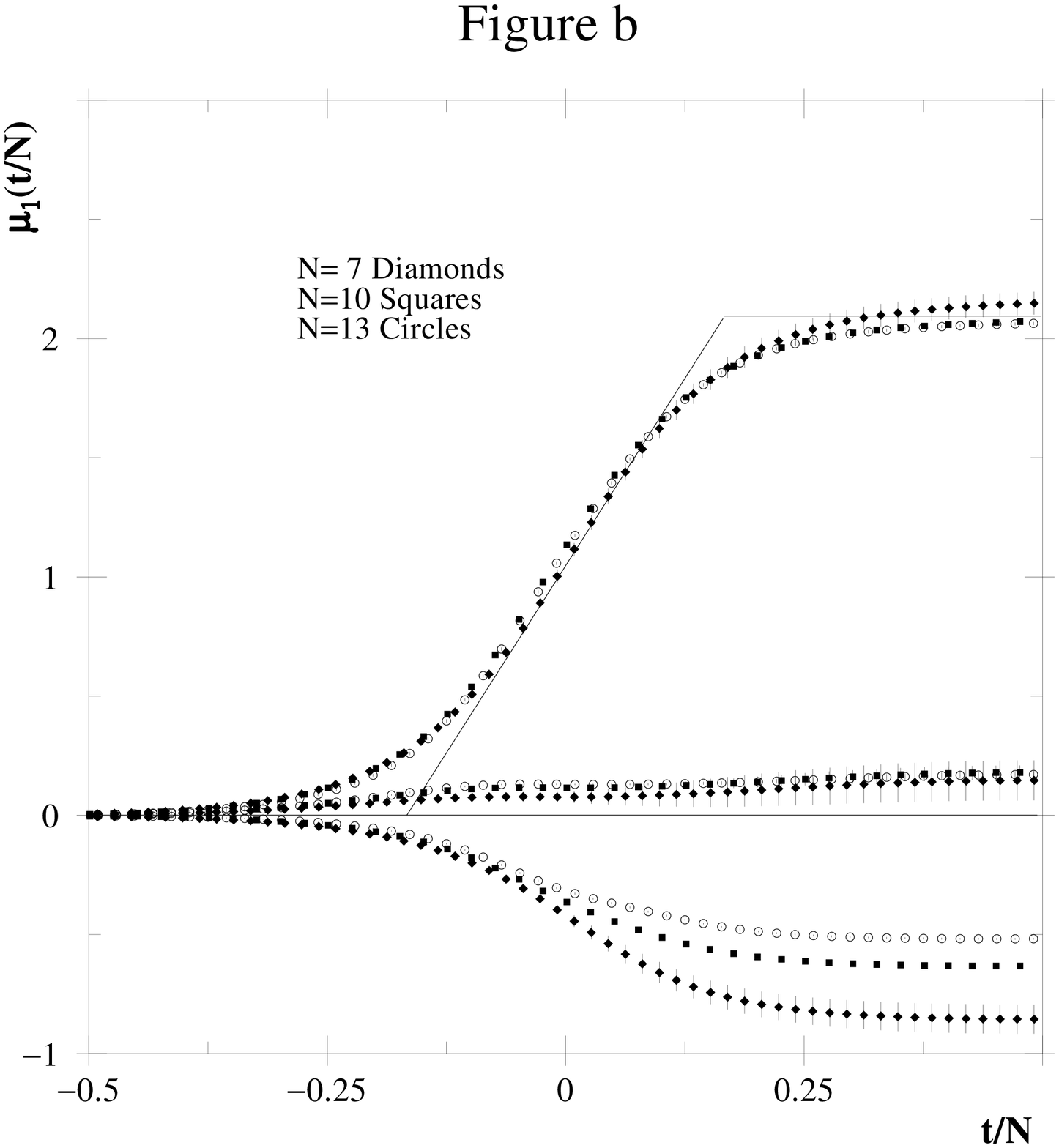   } } }
            }
       \hbox{       \vbox{ \epsfxsize=3truein \hbox{\epsffile{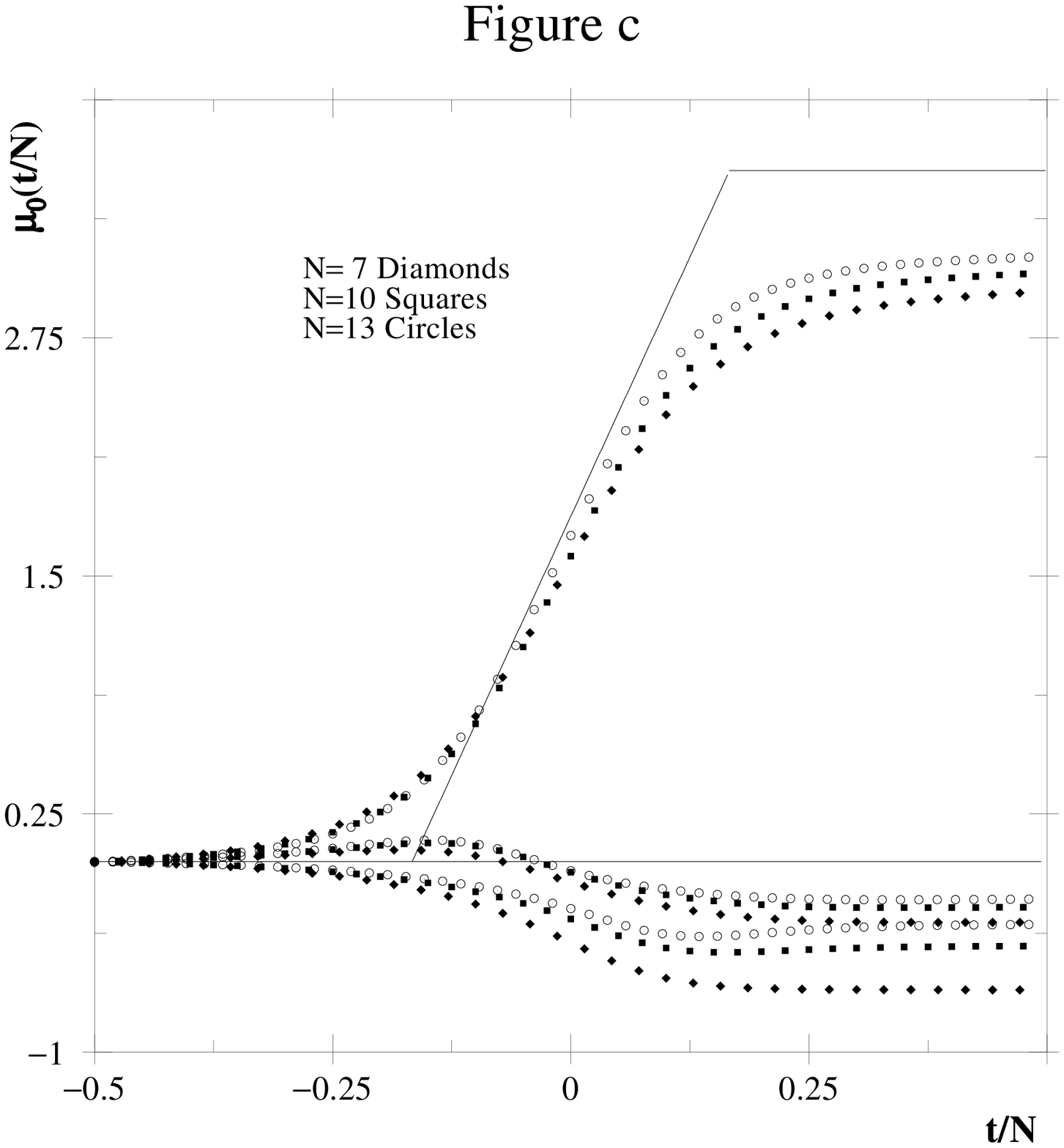   } } }
             \hfill \vbox{ \epsfxsize=3truein \hbox{\epsffile{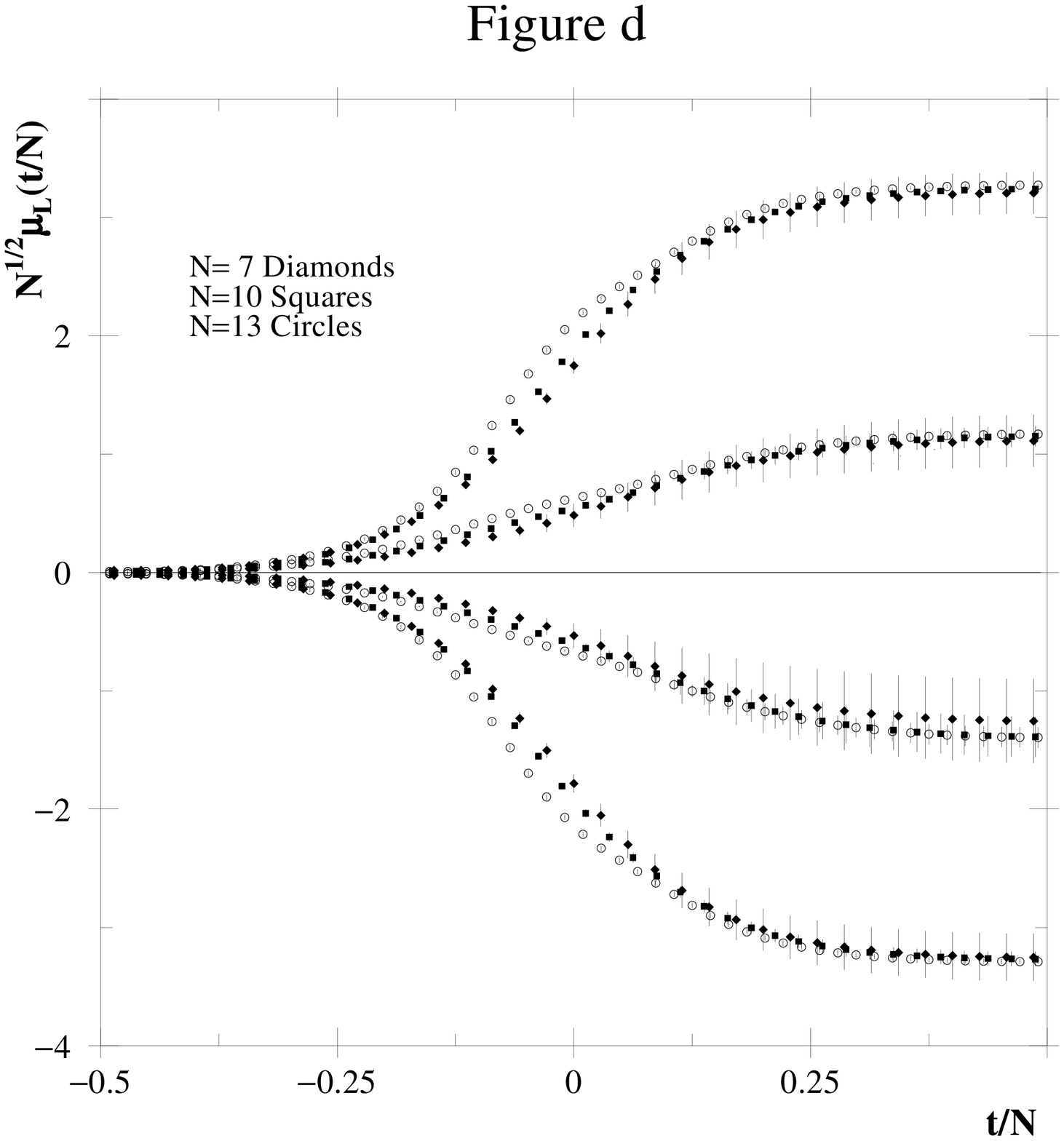   } } }
            }
      }
\caption{{\footnotesize In figure a we show the eigenvalues $\mu_1$ of ${\bf A}_1$ as a
function of time for the solution with $N=7$ and different lattice sizes.  
In figure b it is plotted the eigenvalues $1,2$ and $N$ of ${\bf A}_1$ as a function of $t/N$,
for three different values of $N$.  In figure c the eigenvalues $1,2$ and $N$ of 
${\bf A}_0$ are plotted as a function of $t/N$. In figure d the eigenvalues $1,2,N-1$ and $N$ of ${\bf A}_L$ 
multiplied by $\sqrt{N}$ are plotted as a function of $t/N$ for the same solutions shown in figure b } }
\label{fig:Amu}
\end{figure}

The behavior with N is shown in Figure 4b in which we plot the first, second
and $\N^{th}$ eigenvalues of ${\bf A}_1$ as a function of $t/\N$ 
for $\N=7,10,13$ and $\vec k = (n,n,n)$ (points and error bars have the same 
meaning as before). Very similar results are obtained if 
we plot ${\bf A}_2$ or ${\bf A}_3$ instead of ${\bf A}_1$. 
One of the eigenvalues is approximately independent of N and the
other N-1 become degenerate  and approach zero for increasing $N$. We also 
show the biggest eigenvalue for the abelian selfdual solution described
in section 3 of the appendix in the $N \rightarrow \infty$ limit. In this 
limit the function describing this eigenvalue is $2 \pi(\frac{t}{N}+\frac{1}{6})$
for the values $ - \frac{1}{6} \leq \frac{t}{N} \leq \frac{1}{6}$. We can see that 
the first eigenvalue of $A_1$ is approaching the one of the abelian selfdual solution
in the $N \rightarrow \infty$ limit.

As for ${\bf B}_i$ fields, we consider the following combinations of ${\bf A}_i$ fields,
\bea
\lefteqn{ {\bf A}_0 = \frac{1}{\sqrt{3}} \left(   {\bf A}_1 + {\bf A}_2 + {\bf A}_3  \right)     } \hspace{5.0 cm} \nonumber \\ 
\lefteqn{ {\bf A}_L = \frac{1}{\sqrt{6}} \left( -2{\bf A}_1 + {\bf A}_2 + {\bf A}_3  \right)     } \hspace{5.0 cm} \nonumber \\ 
\lefteqn{ {\bf A}_T = \frac{1}{\sqrt{2}} \left(       - {\bf A}_2 + {\bf A}_3  \right)     } \hspace{4.9 cm} .  
\label{eq:compAbel}
\eea

\noindent
In figure 4c we show the first, second and $N^{th}$ eigenvalues of ${\bf A}_0$ for the 
same solutions appearing in figure 4b (points and error bars have the same meaning 
as before). We observe that the eigenvalue structure is
the same one seen for ${\bf A}_1$ and the expected result if there is a common
component in color space for ${\bf A}_1$, ${\bf A}_2$ and ${\bf A}_3$, the first 
eigenvalue of ${\bf A}_0$ is bigger than the first one for ${\bf A}_i$. We also show
the first eigenvalue of ${\bf A}_0$ for the abelian selfdual solution in the $\N \rightarrow
\infty$ limit, in this case the function $2 \pi \sqrt{3} (\frac{t}{N}+\frac{1}{6})$ for 
points $ -\frac{1}{6} \leq \frac{t}{N} \leq \frac{1}{6} $. The first eigenvalue of $A_0$ is 
approaching the one of the abelian selfdual solution in the $N \rightarrow \infty$ limit.

In figure 4d we show the first, second, $(N-1)^{th}$ and $N^{th}$ eigenvalues of 
$\sqrt{N} \times {\bf A}_L$ for the same solutions appearing in figure 4b (points and error bars 
have the same meaning as before). Very similar results are obtained if we plot the
eigenvalues of ${\bf A}_T$ instead of the ones for ${\bf A}_L$.  The eigenvalue structure is 
completely different to the one shown for ${\bf A}_0$, ${\bf A}_i$. The eigenvalues are 
distributed in pairs, each pair with two opposite values. As can be seen from the figure
these eigenvalues goes to zero as $1/\sqrt{N}$. This means that ${\bf A}_L$ and ${\bf A}_T$
go to zero for large N while ${\bf A}_0$ is independent of $\N$.

{\bf 2. Colour orientation of the potential}. As for the field strength, this is studied by
calculating the commutators of $ {\bf A}_i$'s fields. 
We show in Figure 5 the eigenvalues of $N^{1/2}[{\bf A}_1,{\bf A}_2]$ as a function of $t/\N$ 
for the solutions of Figures 4b, 4c and 4d (again the error bars represent spatial dispersion). 
Only two eigenvalues are relevant being the other N-2 very close to zero and it is also 
clear the scaling with $N^{1/2}$ of the two relevant eigenvalues. From this figure we
conclude that the gauge field  $ {\bf A}_i$ become abelian in the large N limit, compatible
with the properties presented before, $ {\bf A}_0$ is the remaining component in this
limit while $ {\bf A}_L$, ${\bf A}_T$ goes to zero with $N^{1/2}$.
   
\begin{figure}
\caption{{\footnotesize In this figure we plot the eigenvalues $\mu_{12}$ 
of $\imath [{\bf A}_1,{\bf A}_2]$ multiplied by $N^{1/2}$ as a function of $t/N$ for three
values of $N$. } }
\vbox{ \hbox{ \hskip4truecm      \vbox{ \epsfxsize=3truein \hbox{\epsffile{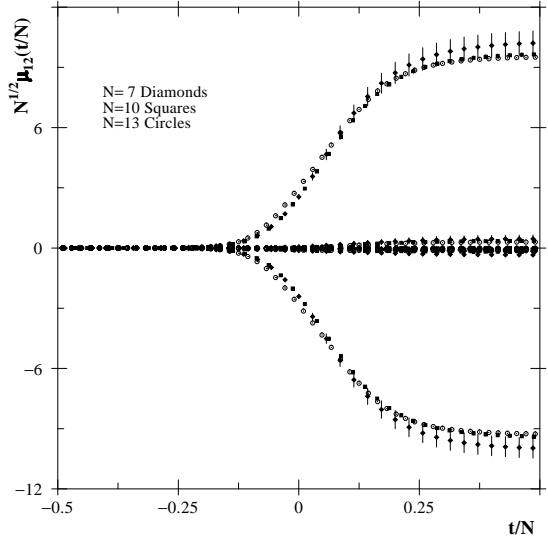} } } } }
\label{fig:noabA}
\end{figure}

{\bf 3. Twist matrix ${\bf \Omega_4}$}.  The temporal twist matrix behaves 
very differently depending of the temporal twist used. Note that their
trace is the temporal Polyakov loop given before.
We study their eigenvalues and obtain the following. For temporal twist
$\vec k=(n,n,n)$ we fit the eigenvalues to the expression: 
\bea
\lefteqn{ \lambda_j = Exp \left( \frac{\imath 2\pi \vec k \vec r} {N(N-1)} + \imath \frac{\pi}{3} 
                                 + \imath N \pi \right) \hspace{1.2 cm}
                            j=1,...,(N-1)/3 } \hspace{14.0 cm} \nonumber \\ 
\lefteqn{ \lambda_j = Exp \left( \frac{\imath 2\pi \vec k \vec r} {N(N-1)} + \imath \pi)
                                 + \imath N \pi \right) \hspace{1.3 cm} 
                            j=(N-1)/3+1,...,2(N-1)/3 } \hspace{14.0 cm}  \nonumber \\ 
\lefteqn{ \lambda_j = Exp \left( \frac{\imath 2\pi \vec k \vec r} {N(N-1)} + \imath \frac{5\pi}{3}
                                 + \imath N \pi \right) \hspace{1.0 cm}
                            j=2(N-1)/3+1,...,N-1 } \hspace{14.0 cm}  \nonumber \\ 
\lefteqn{ \lambda_N = Exp \left( -\frac{i 2\pi \vec k \vec r} {N} +i\pi \right) } \hspace{14.0 cm} 
\eea
This expression was obtained from the numerical data for the eigenvalues of the
twist matrix ${ \bf \Omega_4 } $. This is a good parametrization for all values of
$N$ studied ($N=4,7,10,13,19,25$) and the fits to this expression are better for bigger 
values of N. The interesting point is that this expression for the eigenvalues is
the one for the eigenvalues of the twist matrix ${\bf \Omega_4}$ given in 
\ref{eq:abseO} for the abelian solution described in the appendix.

For temporal twist $\vec k=(1,0,0)$ we obtain for large $N$ 
that the eigenvalues only depend on the $x_1$ coordinate, but we have not found
a good parametrization in this case. 

\section{Conclusions}

In this paper we have presented a set of solutions of the SU(N) Yang Mills 
equations of motion. These solutions are selfdual or antiselfdual, 
have fractional topological charge $Q=1/N$ and live on the four dimensional 
torus, $T^4$. We have studied the case when the lengths of the torus are
$L^3 \times T$ with $T \gg L$ and with twist vectors  $\vec m = (1,1,1)$ and
$\vec k = \frac{N-1}{3} (1,1,1) $ , $\vec k = (1,0,0)$. Now we summarize the main
results we have obtained.

The obtained results show a clear tendency to describe continuum functions,
being the different lattice sizes used enough to observe independence of the
number of lattice points. This property indicates that the obtained configurations
describe continuum Yang-Mills fields. 

For each value of $N$ and twist we always obtain the same solution up to
a gauge transformation and a spatial translation. This means that we can
repeat the procedure to obtain another configuration and the differences 
observed will be a gauge transformation and a spatial translation. 

The main characteristic of the solutions are the following
\begin{itemize} 
\item The obtained solutions are selfdual or antiselfdual in all the studied cases.
      We observe numericaly that this property is satisfied with a very high precision.  
      This guarantees that these configurations are solutions of the equations of 
      motion. 

\item The size of the solutions is approximately $N/3$. By size we understand the length
      of the region in the temporal direction in which the core of the solution is 
      included. We can see this property in all quantities shown in section 3. For 
      example, we can look at the energy profile and check that the most relevant part 
      is located in a region of size $ \sim N/3$. Also in the same region the eigenvalues
      of the field strength take their maximum values and go to zero out of this zone.
      We can say similar assertions for all quantities calculated in this paper.   
      
\item The action density has only one maximum for values of $N \leq 13$ and a double peak
      structure for bigger values. The spatial dependence of the action density disappears
      with increasing $N$ being only dependent of the temporal coordinate.

\item The orientation of the field strength ${\bf F_{\mu \nu} }$ in color space is very 
      dependent on the value of $N$. For smaller values of $N$ the different components
      of ${\bf F_{\mu \nu} }$ are built from different components in color space while 
      for bigger values of $N$ the same component in color space gives the main 
      contribution to the field strength. This property also holds for the gauge field 
      ${\bf A_{\mu}}$. This means that, in the $N \rightarrow \infty$ limit, 
      these solutions are abelians.

\item In the $N \rightarrow \infty$ limit one eigenvalue gives the most important contribution
      to some quantities calculated in this article. In this limit, each component of the field
      strength ${\bf F_{\mu \nu} }$ has one eigenvalue which is approximately $N$ times bigger 
      than the other $N\!-\!1$ eigenvalues. This property also holds for 
      each component of the gauge field ${\bf A_{\mu}}$.  

\item Independence of the temporal twist vector $\vec k$ for some of the calculated quantities. 
      This is an $N$ independent property which holds when the length in the temporal direction
      goes to $\infty$. This property can be seen, for example, in the field strength 
      ${\bf F_{\mu \nu} }$; if we change the temporal twist vectors used, $\vec k_1 =(1,0,0)$
      and $\vec k_2 = (n,n,n)$ we obtain that $\bf B^{\vec k_1}_i = 
      \bf B^{\vec k_2}_i$ and $\bf E^{\vec k_1 }_i = \bf - E^{\vec k_2}_i$. A similar 
      property is held for the gauge field, under the change of the twist vectors we obtain
      the relation ${ \bf A_i^{\vec k_1 }} = - {\bf A_i^{\vec k_2 } }$. 
\end{itemize}
 
We have not succeeded in finding an analytic expression describing the properties of
the studied solutions. Obviously, the first requirement for an ansatz prepared to find the 
analytical expression is that this ansatz satisfies the previously shown properties. The most
promising approach seems to be an ansatz based on the similarity of the solution in the large
$\N$ limit with the abelian solution presented in the appendix. Nevertheless, if the 
solutions in this limit coincide with the abelian solution, something singular must happen
at points $|t/N|=1/6$, being therefore not ease to use this similarity to find the analytic
expression. To conclude, we hope that all the numerical data presented will be helpful for other 
attempts to find the analytical expression of the solutions presented in this paper.

\section*{A. Analytic solutions.}

\noindent
{\bf 't Hooft construction.}
The ${\bf A}_{\mu}$ and ${\bf F}_{\mu \nu}$ fields are built from a 
diagonal matrix ${\bf T}$, in the following way 
\be
{\bf A}_{\mu}(x) = - \frac{\pi}{N}\sum_{\nu} \frac {\alpha_{\mu \nu} x_{\nu}} { l_{\mu} l_{\nu} } {\bf T},
\hspace{1 cm}
{\bf F}_{\mu \nu}(x) =   \frac{2 \pi}{N} \frac {\alpha_{\mu \nu}} { l_{\mu} l_{\nu} }{\bf T}. \label{eq:campos}
\ee
where $\alpha_{\mu \nu}$ is an antisymmetric tensor and $l_{\mu}$ the length of the torus
in the $\mu$ direction. The matrix ${\bf T}$ has the form,   
\be
{\bf T} = \pmatrix{-l \Id_{k \times k}  & {\bf 0}_{k\times l} \cr 
{\bf 0}_{l \times k} & k \Id_{l \times l} }
\ee
being $k$ and $l$ integer numbers ($k+l=N$). To build the twist matrices
we use the ${\bf P}$ and ${\bf Q}$ matrices defined in equation \ref{eq:pandq}. 
From these matrices we construct another set of matrices,
\be
{\bf P}_1, {\bf Q}_1 = \pmatrix{\left( {\bf P}, {\bf Q} \right)_{k \times k}  & {\bf 0}_{k \times l} \cr {\bf 0}_{l \times k} & \Id_{l \times l} }
\hspace{0.3cm} ; \hspace{0.3 cm}
{\bf P}_2, {\bf Q}_2 = \pmatrix{ \Id_{k \times k}  & {\bf 0}_{k \times l} 
\cr {\bf 0}_{l \times k} & \left( {\bf P}, {\bf Q} \right)_{l \times l} }
\ee
satisfying the properties,
\bea
{\bf P}_1 {\bf Q}_1 = {\bf Q}_1 {\bf P}_1 \hspace{0.1 cm} Exp \left\{ \frac{i 2 \pi}{N}
                                      \left(\Id -\frac{{\bf T}}{k}\right) \right\} \hspace{0.3 cm}
                                      ; \hspace{0.3 cm}       
{\bf P}_2 {\bf Q}_2 = {\bf Q}_2 {\bf P}_2 \hspace{0.1 cm} Exp \left\{ \frac{i 2 \pi}{N}
                                      \left(\Id +\frac{{\bf T}}{l}\right) \right\} . \label{eq:conm}
\eea
And the ansatz for the twist matrices is,
\be
{\bf \Omega}_{\mu}(x) = {\bf P}_1^{s_{\mu}} {\bf Q}_1^{t_{\mu} } {\bf P}_2^{u_{\mu}} {\bf Q}_2^{v_{\mu}} \hspace{0.1 cm}
                  Exp \left\{ - \frac{i \pi}{N} \sum_{\nu} \frac {\alpha_{\mu \nu} x_{\nu}}
                                                       { l_{\nu} } {\bf T} \right\} \label{eq:mattwi}
\ee
where $s_{\mu}$,$t_{\mu}$,$u_{\mu}$ and $v_{\mu}$ are arbitrary integer numbers. 
These matrices must satisfy the consistency condition,
\bea
{\bf \Omega}_{\mu}(x_{\nu} + {l_{\nu}} ) {\bf \Omega}_{\nu}(x_{\mu} ) 
= {\bf \Omega}_{\nu}(x_{\mu} + {l_{\mu}} ) {\bf \Omega}_{\mu}(x_{\nu} ) 
 \ {\rm Exp} \left(-\frac{i 2 \pi n_{\mu \nu} }{\N} \right) . \label{eq:concon}
\eea
This condition imposes the following equations for $s_{\mu}$,$t_{\mu}$,$u_{\mu}$ and $v_{\mu}$
\bea
\frac{1}{k}( t_{\mu} s_{\nu} - t_{\nu} s_{\mu}) =
   l \frac{\alpha_{\mu \nu}}{N} + \frac{n_{\mu \nu}}{N} + A_{\mu\nu} \hspace{0.3 cm} ;   \hspace{0.3 cm}
\frac{1}{l}( v_{\mu} u_{\nu} - v_{\nu} u_{\mu}) =
  -k \frac{\alpha_{\mu \nu}}{N} + \frac{n_{\mu \nu}}{N} + B_{\mu\nu} \label{eq:twoeq}
\eea 
where $A_{\mu\nu}$ and $B_{\mu\nu}$ are integer numbers. 
To solve these two equations we give to $\alpha_{\mu\nu}$ and $n_{\mu\nu}$ the form,
\bea
n_{\mu\nu}=n_{\mu\nu}^{(1)}+n_{\mu\nu}^{(2)}  \hspace{1 cm} ; \hspace{1 cm} 
\alpha_{\mu\nu}=\frac{n_{\mu\nu}^{(1)}}{k}-\frac{n_{\mu\nu}^{(2)}}{l} \label{eq:tensal}
\eea
and then equations in formula \ref{eq:twoeq} are transformed to,
\bea
n_{\mu \nu}^{(1)}= t_{\mu} s_{\nu} - t_{\nu} s_{\mu} + kA_{\mu \nu} \hspace{0.5 cm} ; \hspace{0.5 cm} 
n_{\mu \nu}^{(2)}= v_{\mu} u_{\nu} - v_{\nu} u_{\mu} + lB_{\mu \nu} \label{eq:intsol}
\eea
which can be solved if the following condition is satisfied,
\bea
n_{\mu \nu}^{(1)} \tilde{n}_{\mu \nu}^{(1)} = 0
 \hspace{0.5 cm} mod \hspace{0.1 cm} k \hspace{0.8 cm} ; \hspace{0.8 cm}
n_{\mu \nu}^{(2)} \tilde{n}_{\mu \nu}^{(2)} = 0 \hspace{0.5 cm} mod \hspace{0.1 cm} l 
\eea
The $n_{\mu \nu}^{(1)}$ and $n_{\mu \nu}^{(2)}$ tensors are orthogonal twist tensors
for a SU(k) and SU(l) group respectively.

The topological charge for these solutions is,
\be
Q =  \frac{1}{N} \frac{\alpha_{\mu \nu} \tilde{\alpha}_{\mu \nu}}{4} k l   
  =  \frac{1}{N} \left( \vec k^{(1)}\vec m^{(1)} \frac{l}{k} + 
                        \vec k^{(2)}\vec m^{(2)} \frac{k}{l} -
                        \vec k^{(1)}\vec m^{(2)} - \vec k^{(2)}\vec m^{(1)} \right) \label{eq:topcha}
\ee
where we have defined the vectors ${k_i}^{(n)}={n_{0i}}^{(n)}$,
${m_i}^{(n)}=\epsilon_{ijk}{n_{jk}}^{(n)}/2$, with $n=1,2$.  

\vspace{0.2 cm}

\noindent
{\bf Some examples.}
Now we give some examples of solutions built using 't Hooft construction, and for the
torus lengths used in this article: $l_x=l_y=l_z=1$ and $l_t \rightarrow \infty$. In fact,
our examples will be given for any value of $l_t$. The topological charge in our examples
is given by equation \ref{eq:topcha} and the action by the following equation,  
\bea
S=\frac{8\pi^2}{N} \frac{1}{2}
                     \left\{ \frac{1}{l_t} \left( \vec k^{(1)}\vec k^{(1)} \frac{l}{k} +
                                                \vec k^{(2)}\vec k^{(2)} \frac{k}{l} - 
                                                2 \vec k^{(1)}\vec k^{(2)} 
                                         \right) +  \right. \hspace{0.7 cm} \nonumber \\ 
                      \left.              l_t  \left( \vec m^{(1)}\vec m^{(1)} \frac{l}{k} +
                                                \vec m^{(2)}\vec m^{(2)} \frac{k}{l} - 
                                                2 \vec m^{(1)}\vec m^{(2)} 
                                         \right) 
                   \right\}  \hspace{-0.1 cm}  .
\label{eq:accar}
\eea
The minimum value for the action, $S=8\pi^2|Q|$, is obtained when the solution 
is selfdual or antiselfdual. This condition imposes a value for the temporal length
$l_t$, obtained solving the equation,
\be
\frac{\vec k^{(1)}}{l} - \frac{\vec k^{(2)}}{k} =
   \pm l_t \left( \frac{\vec m^{(1)}}{l} - \frac{\vec m^{(2)}}{k} \right) \label{eq:selfeq}
\ee
the positive sign for selfdual solutions and the negative sign for antiselfdual solutions.
Now we give some examples for the twist vectors used in this article:
\begin{enumerate}
\item Solutions for twist  vectors $\vec m=(1,1,1)$ and $\vec k=(n,n,n)$, with
$N=3n+1$ for $n=1,2,3,...$. We choose $k=N -1$ and $l=1$ and the twist vectors in
subspaces SU(k) and SU(l) as,
\bea
\vec m^{(1)}=(1,1,1)   \hspace{0.5 cm }  \vec k^{(1)}=(n,n,n) 
\hspace{1 cm} ; \hspace{1 cm}  
\vec m^{(2)}=(0,0,0)   \hspace{0.5 cm }  \vec k^{(2)}=(0,0,0)   \nonumber 
\eea
The topological charge for this solution is $Q=1/N$ and it is selfdual when
$l_t = n$.

\item Solutions for twist vectors $\vec m=(1,1,1)$ and $\vec k=(1,0,0)$.
We choose the twist vectors in $SU(k)$ and $SU(l)$ subspaces as,
\bea
\vec m^{(1)}=(1,1,1) \hspace{0.5 cm }  \vec k^{(1)}=(0,0,0)  
\hspace{1 cm} ; \hspace{1 cm}
\vec m^{(2)}=(0,0,0) \hspace{0.5 cm }  \vec k^{(2)}=(1,0,0)  \nonumber
\eea
This choice works for any values of $k,l$. The topological charge for this solution
is $Q=-1/N$ and in this case it is not possible to solve the selfduality equation.
\end{enumerate}

\vspace{0.2 cm}

\noindent
{\bf  Changing the gauge.}
Now we change the gauge for one of the solutions described before to the gauge 
used in section 3. We choose the solution given in example 1. In this case
$k=N-1$, $l=1$, $n_{\mu\nu}^{(1)}=n_{\mu\nu}$ and $n_{\mu\nu}^{(2)}=0$.
The fields are, 
\bea
{\bf A}_{\mu}(x) = - \frac{\pi}{N(N-1)}\sum_{\nu} \frac {n_{\mu \nu} x_{\nu}}
                                             { l_{\mu} l_{\nu} } {\bf T},
\hspace{1 cm}
{\bf F}_{\mu \nu}(x) =   \frac{2 \pi}{N(N-1)} \frac {n_{\mu \nu}}
                          { l_{\mu} l_{\nu} }{\bf T}. \label{eq:abseF}
\eea
we remember that the torus lengths were $l_x=l_y=l_z=1$ and  $l_t$ can take any value, and
the twist vectors were  $\vec m = (1,1,1)$ and $\vec k = (n,n,n)$, with  $N=3n+1$.
The action and the topological charge take the values,
\bea
  S = \frac{8 \pi^2}{N} \frac{1}{2} \left( \frac{l_t}{n}+ \frac{n}{l_t} \right) \hspace{0.3 in} Q = \frac{1}{N}  \nonumber
\eea
the twist matrices are,
\be
{\bf \Omega}_{\mu}(x) = {\bf P}_1^{s_{\mu}} {\bf Q}_1^{t_{\mu} } \hspace{0.1 cm}
                  Exp \left\{ - \frac{i \pi}{N(N-1)} \sum_{\nu} \frac {n_{\mu \nu} x_{\nu}}
                               { l_{\nu} } {\bf T} \right\} \nonumber
\ee
and the values for $s_{\mu}$, $t_{\mu}$, 
\bea
s_{1} = 0  \hspace{.2 in} s_{2} = 1 \hspace{.2 in} s_{3} = N-2 \hspace{.2 in}
                                                   s_{4} = 2(N-1)/3 \nonumber \\
t_{1} = 1  \hspace{.2 in} t_{2} = 0 \hspace{.2 in} t_{3} = N-2 \hspace{.2 in}
                                                   t_{4} =  (N-1)/3 \hspace{0.2 cm} .\nonumber
\eea
The gauge used in section 3 is: ${\bf A}_4=0$ and ${\bf A}_i(t=-\infty)=0$. For our example we can not use
the same gauge because  $t$ is finite and ${\bf F}_{\mu\nu} \neq 0$ for any value of $t$ (this is the
condition needed to put $A_i(t)=0$ at some point $t$). The most similar gauge condition is the
following one,
\bea
{\bf A}_4=0 , \hspace{0.2 in}
{\bf A}_3(t=0)=0 , \hspace{0.2 in}
{\bf A}_2(t=0,z=0)=0 , \hspace{0.2 in}
{\bf A}_1(t=0,z=0,y=0)=0 \nonumber
\eea
because all links associated with these fields were rotated to the identity.
The field ${\bf A}_{\mu}$ and the twist matrices in this gauge are,
\bea
\lefteqn{ {\bf A}_4 = 0 } \hspace{9 cm}  \nonumber \\
\lefteqn{ {\bf A}_3 = \frac{2\pi}{N(N-1)} {\bf T} \hspace{0.05 in} \frac{n}{l_t}t  }
     \hspace{9 cm} \nonumber \\
\lefteqn{ {\bf A}_2 = \frac{2\pi}{N(N-1)} {\bf T} \hspace{0.05 in} \left( \frac{n}{l_t}t-z \right) }
 \hspace{9 cm} \nonumber \\
\lefteqn{ {\bf A}_1 = \frac{2\pi}{N(N-1)} {\bf T} \hspace{0.05 in} \left( \frac{n}{l_t}t+z-y \right) }
    \hspace{9 cm}  \label{eq:abseA} \\
\lefteqn{ {\bf \Omega}_4 = {{\bf P}_1}^{s_4}{ {\bf Q}_1}^{t_4}Exp \left\{ - i \frac{2\pi}{N(N-1)} {\bf T}
                           \hspace{0.2 cm} \vec k \vec r \right\}  }\hspace{9 cm} \nonumber \\
\lefteqn{ {\bf \Omega}_3 = {{\bf P}_1}^{s_3}{{\bf Q}_1}^{t_3}Exp \left\{ i\frac{2\pi}{N(N-1)} {\bf T}
             \hspace{0.2 cm}    (y-x) \right\}  } \hspace{9 cm} \nonumber \\
\lefteqn{ {\bf \Omega}_2 = {{\bf P}_1}^{s_2}{{\bf Q}_1}^{t_2}Exp \left\{ i \frac{2\pi}{N(N-1)} {\bf T}
             \hspace{0.2 cm}     x \right\} } \hspace{9 cm}\nonumber \\
\lefteqn{ {\bf \Omega}_1 = {{\bf P}_1}^{s_1}{{\bf Q}_1}^{t_1} } \hspace{9 cm} \label{eq:abseO}
\eea
We compare along section 3 the results obtained for large values of N with this
solution (with $l_t=n$) because some properties are very similar for both solutions in this limit.

Finally, we give the value of the Polyakov loops for this example,
\bea
{\cal L}_{\mu} = \frac{1}{N} Exp \left( - i \frac{2 \pi}{N} \sum_{\nu} \frac{n_{\mu\nu} x_{\nu}}{l_{\nu}} 
 \right) \nonumber
\eea 
note that this quantity is gauge invariant and could be calculated with the two
different ${\bf A}_{\mu}$ given before, obtaining the same result.

\section*{Acknowledgements}
The results presented in this article are part of the Ph. D. Thesis of the author \cite{tesis}. 
Most of the ideas developed in this work come from suggestions and fruitful discussions with 
Antonio Gonz\'alez-Arroyo. I also acknowledge useful discussions with Margarita Garc\'{\i}a
P\'erez and Carlos Pena. This work has been supported by the Spanish Ministerio de Educaci\'on 
y Cultura under a postdoctoral Fellowship and by CICYT under grant AEN97-1678.

\end{document}